\documentclass[pre,aps]{revtex4-1}
\usepackage{graphicx}% Include figure files
\usepackage{bm}% bold math
\usepackage{tabularx}
\usepackage{amsmath}

\begin{document}

\title{Dynamical tunneling and control}
\author{Srihari Keshavamurthy}
\affiliation{Department of Chemistry, Indian Institute of Technology, Kanpur, 
Uttar Pradesh 208016, India}

\begin{abstract}
Contribution to the edited volume {\em Dynamical Tunneling: Theory and Experiment}, Editors Srihari Keshavamurthy and Peter Schlagheck, CRC Press , Taylor \& Francis, Boca Raton, 2011.
\end{abstract}
\maketitle

\section{Introduction}

Tunneling is a supreme quantum effect. Every introductory text\cite{introqmbook} on quantum mechanics gives the
paradigm example of a particle tunneling through a one-dimensional potential barrier despite
having a total energy less than the barrier height. Indeed, the reader typically works through
a number of excercises, all involving one-dimensional potential barriers of
one form or another modelling several key physical phenomena ranging from
atom transfer reactions to the decay of alpha particles\cite{merz02}. However,
one seldom encounters coupled multidimenisonal tunneling in such texts since an analytical solution
of the Schr\"{o}dinger equation in such cases is not possible. Interestingly,
the richness and complexity of the tunneling phenomenon manifest themselves in full glory in the
case of multidimensional systems\cite{multidtun}. 
Thus, for instance, the usual one-dimensional expectation
of increasing tunneling splittings as one approaches the barrier top from below is not
necessarily true as soon as one couples another bound degree of freedom to the tunneling coordinate.
In the context of molecular reaction dynamics, multidimensional tunneling can result in
strong mode-specificity and fluctuations in the reaction rates\cite{millerpaps}. 
In fact, a proper description of tunneling
of electrons and hydrogen atoms is absolutely essential\cite{largemoltun,bo2000} even in molecular systems as large as
enzymes and proteins. 
Although one usually assumes tunneling effects to be significant
in molecules involving light atom transfers it is worth pointing out that neglecting
the tunneling of even a heavy atom like carbon is the difference between a reaction
occuring or not occuring. In particular, one can underestimate rates by
nearly hundred orders of magnitude\cite{catomtun}. Interestingly, and perhaps paradoxically,
several penetrating insights into the nature and mechanism of
multidimensional barrier tunneling have been obtained from a phase space perspective\cite{creagh,shudo}.
The contributions by Creagh, Shudo and Ikeda, and Takahashi in the present volume
provide a detailed account of the latest advances in the phase space based understanding
of multidimensional barrier tunneling.

What happens if there are no coordinate space barriers? In other words, in situations wherein
there are no static energetic barriers separating ``reactants" from the ``products" 
does one still have to be concerned about quantum tunneling?
One such model potential is shown in Fig.~\ref{FIG1} which will be discussed in the next section.
Here we have the notion of reactants and products in a very general sense. So, for instance,
in the context of a conformational reaction they might correspond to the
several near-degenerate conformations of a specific molecule. 
Naively one might expect
that tunneling has no consequences in such cases. However, 
studies over last several decades\cite{lc79,dh81,hp84,Ozo84,srh82c,sp87,gh87,lb90,btu93,tu94,df95} have
revealed that things are not so straightforward. Despite the lack of static barriers, the 
dynamics of the system can generate barriers and quantum tunneling can occur through such
dynamical barriers\cite{dtunbook1}. This, of course, immediately implies that dynamical tunneling is a very
rich and subtle phenomenon since the nature and number of barriers can vary appreciably 
with changes in the nature of the dynamics over
the timescales of interest. This would also seem to imply that deciphering the mechanism
of dynamical tunneling is a hopeless task as opposed to the static potential barrier
case wherein elegant approximations to the tunneling rate and splittings can be written down.
However, recent studies have clearly established that even in the case of
dynamical tunneling it is possible to obtain very accurate approximations to the splittings
and rate. In particular, it is now clear that unambiguous identification
of the local dynamical barriers is possible only by a detailed study of the structure
of the underlying classical phase space. The general picture that has emerged is that
dynamical tunneling connects two or more classically disconnected regions in the phase space.
More importantly, and perhaps ironically, the dynamical tunneling splittings and rates
are extremely sensitive to the various phase space structures like 
nonlinear resonances\cite{bsu01,bsu02,seu05,es05,bfgr98,ablr01,rbiwg94,sfgr06},
chaos\cite{gh87,lb90,btu93,tu94,Ort96,ammk95} and partial barriers\cite{grr86,pre00-mh-chap3}. 
It is crucial to note that although purely quantum
approaches can be formulated for obtaining the tunneling splittings, any mechanistic
understanding requires a detailed understanding of the phase space topology. In this sense,
the phenomenon of dynamical tunneling gets intimately linked to issues related to
phase space transport. Thus, one now has the concept of resonance-assisted tunneling (RAT)
and chaos-assisted tunneling (CAT) and realistic systems typically involve both
the mechanisms.

Since the appearance of the first book\cite{dtunbook1} on the topic of interest
more than a decade ago, there have been several
beautiful experimental studies\cite{ns97,hadghhrr05,hetal01,sor01,bafmli03,shfhsn10} that have revealed various aspects of the
phenomenon of dynamical tunneling. The most recent one by Chaudhury {\it et al.}
realizes\cite{csagj09} the paradigmatic kicked top model using cold $^{133}$Cs atoms and 
clearly demonstrate the dynamical tunneling occuring in the underlying phase space.
Interestingly, good correspondence between the quantum dynamics and classical
phase space structures is found despite the system being in a deep quantum regime.
As another example, I mention the experimental observation\cite{ftcfswmb07} 
by F\"{o}lling {\it et al.} of second order
co-tunneling of interacting ultracold rubidium atoms in a double well trap.
The similarities between this system and the studies on dynamical tunneling
using molecular effective Hamiltonians is striking. In particular, the description of
the cold atom study in terms of superexchange (qualitative and quantitative)
is reminiscent of the early work by Stuchebrukhov and Marcus\cite{smdyntun93} on understanding the
role of dynamical tunneling in the phenomenon of intramolecular energy flow.
Further details on the experimental realizations can be found in this volume
in the articles by Steck and Raizen, and Hensinger. An earlier review\cite{Kes07} provides 
extensive references to the experimental manifestations of dynamical tunneling
in molecular systems in terms of spectroscopic signatures. Undoubtedly, in the coming
years, one can expect several other experimental studies which will lead to
a deeper understanding of dynamical tunneling and raise many intriguing issues
related to the subject of classical-quantum correspondence.

As remarked earlier, it seems ironic that a pure quantum effect like tunneling should bear
the marks of the underlying classical phase space structures. However, it is useful to
to recall the statement by Heller that tunneling is only meaningful with
classical dynamics as the baseline. Thus, insights into the nature of the classical dynamics
translates into a deeper mechanistic insight into the corresponding quantum dynamics.
Indeed, one way of thinking about classical-quantum correspondence is that
classical mechanics is providing us with the best possible ``basis" to describe the
quantum evolution. The wide range of contributions in this volume are a testimony to
the richness of the phenomenon of dynamical tunneling and the utility of 
such a classical-quantum correspondence perspective. 
In this article I focus on the specific field of
quantum control and show as to how dynamical tunneling can lead to useful insights
into the control mechanism\cite{note0,asthesis}. 
The hope is that more such studies will eventually
result in control strategies which are firmly rooted in the intutive classical
world, yet accounting for the classically forbidden pathways and mechanisms
in a consistent fashion. This is a tall order, and some may even argue as an
unnecessary effort in these days of fast computers and smart and efficient 
algorithms to solve fairly high dimensional quantum dynamics.
However, in this context, it is useful to remember the following which was written by
Born, Heisenberg, and Jordan nearly eighty years ago\cite{bhj}

``{\em The starting point of our theoretical approach was the conviction that the
difficulties that have been encountered at every step in quantum theory in the last few years
could be surmounted only by establishing a mathematical system for the mechanics of atomic
and electronic motions, which would have a unity and simplicity comparable with the system of
classical mechanics...further developement of the theory, an important task will lie in the
closer investigation of the nature of this correspondence and in the description of the manner in which
symbolic quantum geometry goes over into visualizable classical geometry.}"

The above remark was made in an era when computers were nonexistent. Nevertheless, it is 
remarkably prescient since even in the present era one realizes the sheer difficulty in implementing
an all-quantum dynamical study on even relatively small molecules\cite{lt10}. 
In any case, it is not
entirely unreasonable to argue that large scale
quantum dynamical studies will still require some form of an implicit
classical-quantum correspondence approach to grasp the underlying mechanistic details.
With the above remark in mind I start things off by revisiting the original paper\cite{dh81} by Davis and Heller
since, in my opinion, it is ideal from the pedagogical point of view.

\section{Davis-Heller system revisited}

Three decades ago, Davis and Heller in their pioneering study\cite{dh81} gave a clear example of dynamical tunneling.
A short recount of this work including the famous plots of the classical trajectories and the
associated quantum eigenstates can be found in Heller's article in this volume. However, I revisit this
model here in order to bring forth a couple of important points that seem to have been
overlooked in subsequent works. First, the existence of another class of eigenstate pairs, called
as circulating states\cite{dh81}, which can exert considerable influence on the usual tunneling doublets at
higher energies. Second, a remark
in the original paper\cite{dh81} which can be considered as a harbinger for chaos-assisted tunneling. As shown below,
there are features in the original model that are worth studying in some detail even after three decades
since the original paper was published. 

\begin{figure}[t]
\includegraphics*[height=120mm,width=150mm]{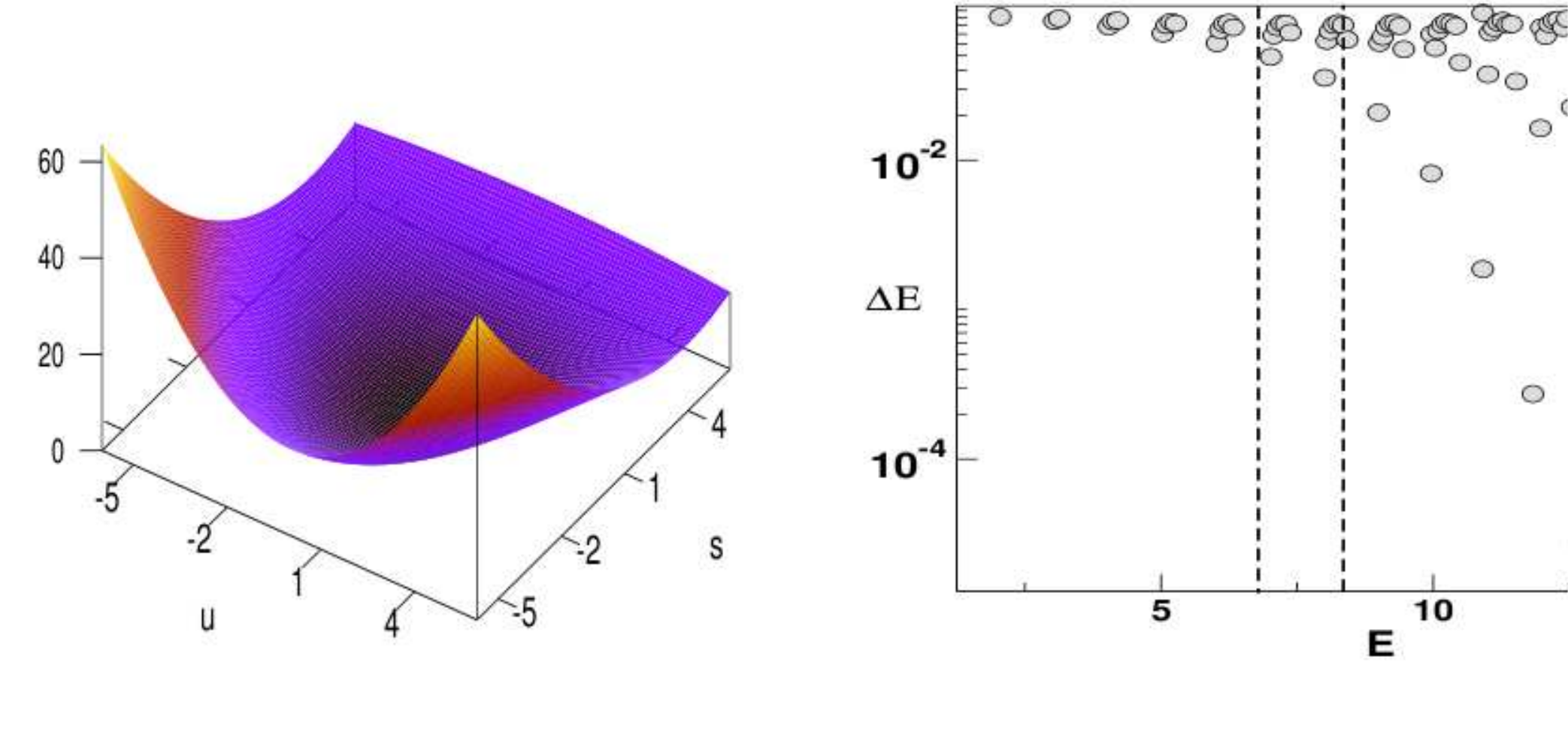}
\caption{The left plot shows the two-dimensional potential used by Davis and Heller to illustrate
the phenomenon of dynamical tunneling. Note that there are no barriers in the coordinate space.
On the right the magnitude of the splitting between adjacent eigenstates are shown as a function
of the eigenenergies. The first and second dotted vertical lines indicate the energies around
which the symmetric (s) and unsymmetric (u) stretch modes become unstable. Sequences of
doublets with very small splittings can be seen around the onset of the instabilities. On the
other hand, the regularity of the sequences vanishes at higher energies.}
\label{FIG1}
\end{figure}

The Hamiltonian of choice is the two degrees of freedom (2DoF, in what follows
the acronymn DoF stands for degrees of freedom) Barbanis-like model\cite{dh81}
\begin{equation}
H(s,u,p_{s},p_{u}) = \frac{1}{2}\left(p_{s}^{2}+p_{u}^{2}\right) + 
                     \frac{1}{2}\left(\omega_{s}^{2} s^{2} + \omega_{u}^{2} u^{2} \right)
                     + \lambda s u^{2}
\end{equation}
with the labels `s' and `u' denoting the symmetric and unsymmetric stretch modes respectively. 
The above Hamiltonian has also been studied\cite{hsd80} in great detail to uncover the correspondence between
classical stability of the motion and quantum spectral features, wavefunctions, and
energy transfer.
The potential
is symmetric with respect to $u \leftrightarrow -u$ as shown in Fig.~\ref{FIG1} but there is
no potential barrier. Davis and Heller used the parameter values $\omega_{s}=1.0,\omega_{u}=1.1$,
and $\lambda=-0.11$ for which the dissociation energy 
$E_{\rm dis} \equiv \omega_{s}^{2}\omega_{u}^{4}/8\lambda^{2} = 15.125$. Note that the masses
are taken to be unity and one is working in units such that $\hbar=1$. The key observation by
Davis and Heller was that despite the lack of any potential barriers several bound eigenstates
came in symmetric-antisymmetric pairs $|\psi_{1}\rangle$ and $|\psi_{2} \rangle$ with energy splittings much smaller than
the fundamental frequencies {\it i.e.,} $\Delta E \equiv |E_{1}-E_{2}| \ll O(1)$.
In Fig.~\ref{FIG1} the various splittings between adjacent eigenstates are shown and it is clear that
several ``tunneling" pairs appear above a certain threshold energy.

\begin{figure}
\begin{center}
\includegraphics*[height=100mm,width=150mm]{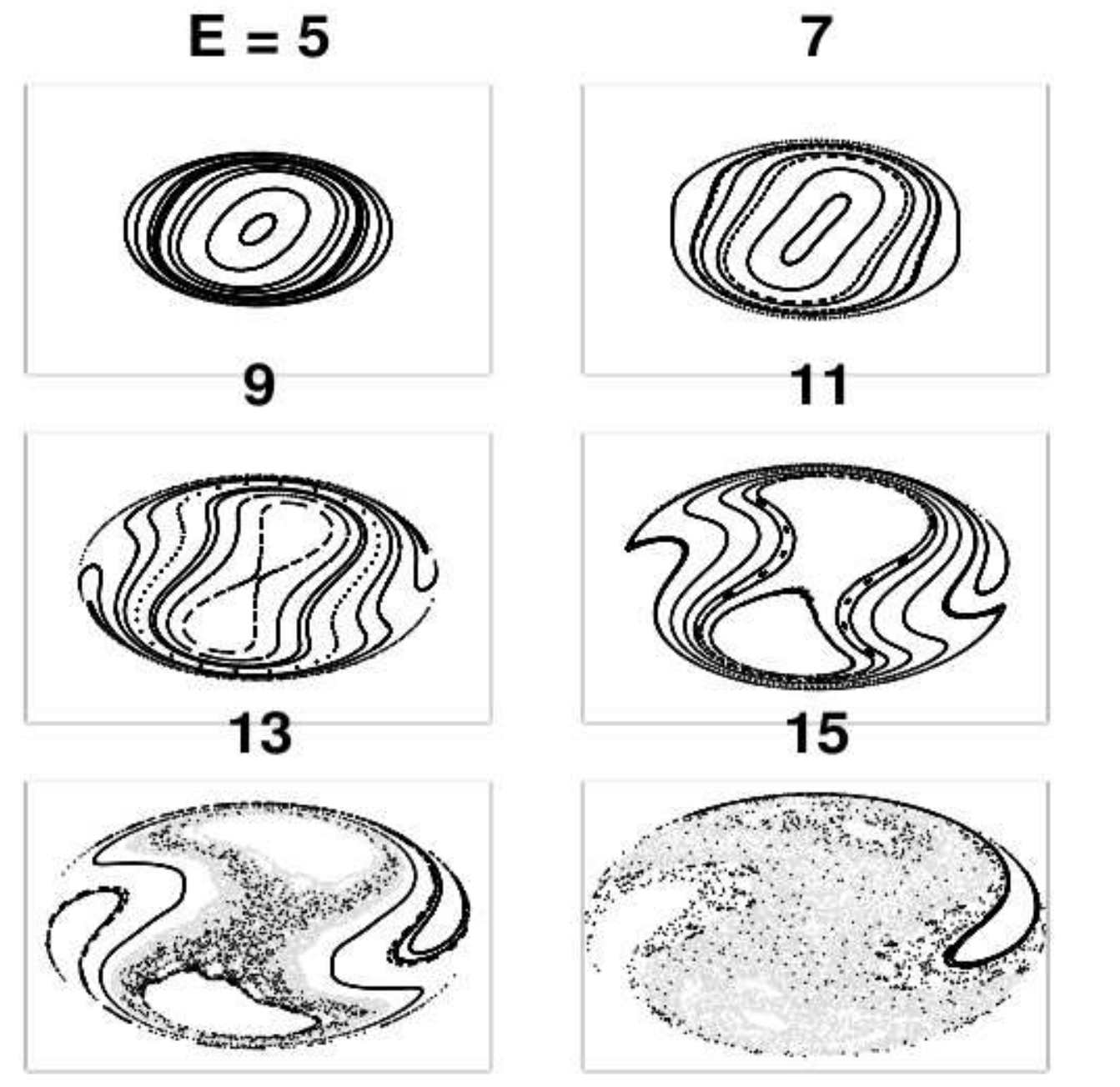}
\caption{Evolution of the classical phase space with increasing energy shown
as $(u,p_{u})$ Poincar\'{e} surface of sections. Note that large scale chaos appears
for $E \geq 11$. The formation of a separatrix and two classically disjoint regular regions can be seen
at $E=9$ due to the symmetric mode periodic orbit becoming unstable. The regular regions
almost vanish near the dissociation energy. In the bottom row some of the chaotic orbits
have been suppressed, for clarity of the figure, by showing them in gray.}
\label{FIG2}
\end{center}
\end{figure}

How can one understand the onset of such near degeneracies in the system? The crucial insight
that Davis and Heller provided was that
the appearance of such doublets is correlated with
the large scale changes in the classical phase space. The nature of the phase space
with increasing total energy is shown in Fig.~\ref{FIG2} using the $(u,p_{u})$ Poincar\'{e}
surface of section. Such a surface of section, following standard methods, is constructed
by recording the points, with momentum $p_{s} > 0$, of the intersection of a trajectory with the
the $s=0$ plane in the phase space. Such a procedure for several trajectories with 
specific total energy generates
a typical surface of section as shown in Fig.~\ref{FIG2} and clearly indicates the
global nature of the phase space.
It is clear from the figure that the phase space for $E < 11$ is mostly
regular while higher energy phase spaces exhibit mixed regular-chaotic dynamics. 
One of the most prominent change happens when the
symmetric stretch periodic orbit $(u=0,p_{u}=0)$ becomes unstable around $E \approx 6.8$. In Fig.~\ref{FIG2}
the consequence of such a bifurcation can be clearly seen for $E=9$ as the formation
of two classically disjoint regular regions. In fact, the two regular regions signal a $1$:$1$ resonance
between the stretching modes. The crucial point to observe here is that
classical trajectories initiated in one of the regular regions cannot
evolve into the other regular region. With increasing energy the classically disjoint regular regions move
further apart and almost vanish near the dissociation energy. The result presented in Fig.~\ref{FIG1}
in fact closely mirrors the topological changes, shown in Fig.~\ref{FIG2}, in the phase space. Thus, 
in Fig.~\ref{FIG1} a sequence of eigenstates with very small splittings begins right
around the energy at which the symmetric periodic orbit becomes unstable. The
unsymmetric stretch, however, becomes unstable at a higher energy $E \approx 8.3$ and one can observe
in Fig.~\ref{FIG1} another sequence that seemingly begins near this point. Note that at higher energies
it is not easy to identify any more sequences, but small splittings are still observed. The important 
thing to note here is that not only do the splittings 
but the individual eigenstates also correlate with the changes in the phase space.

\begin{figure}[t]
\includegraphics*[height=120mm,width=160mm]{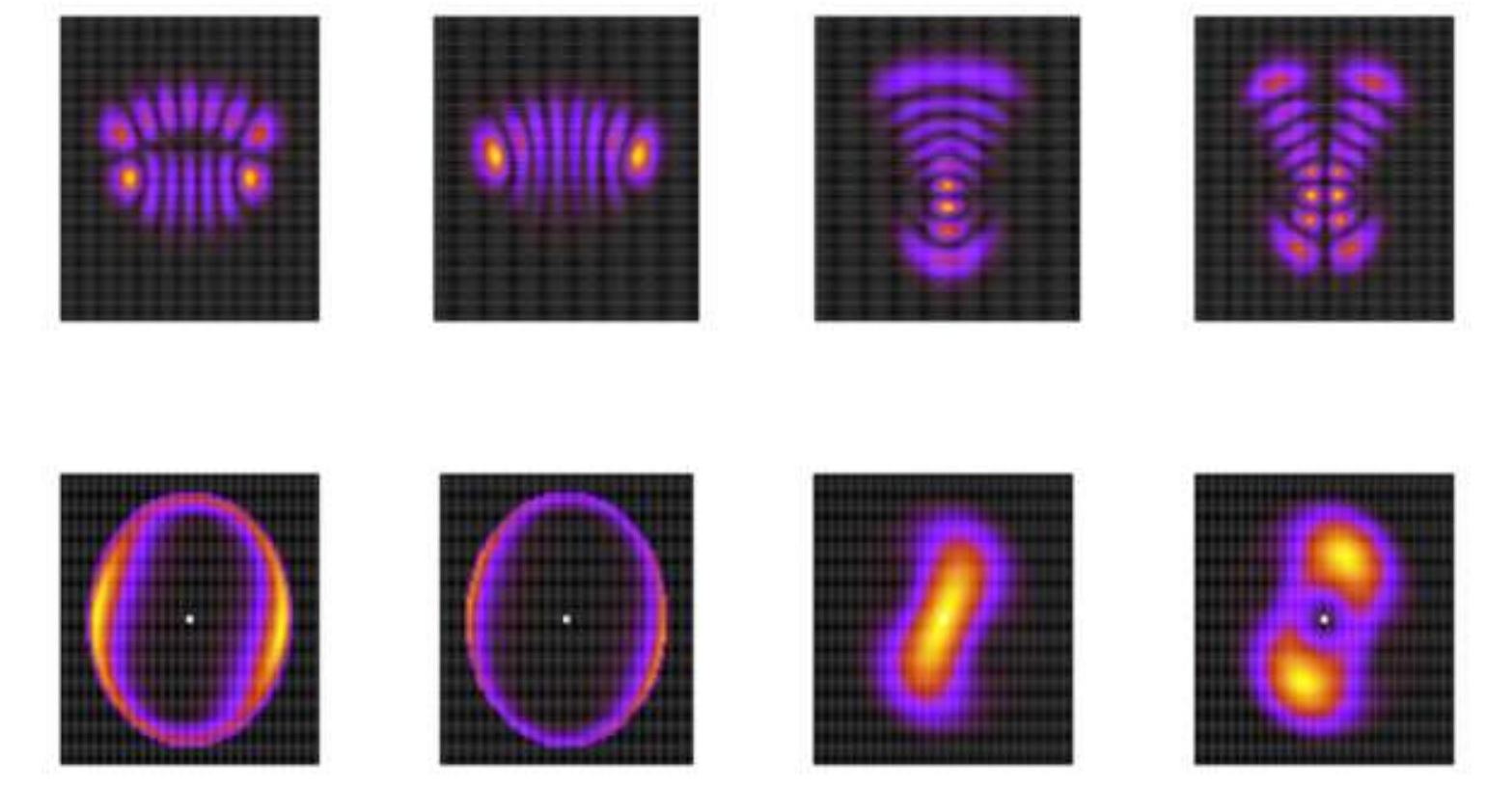}
\caption{Four eigenstates close to an energy where the $1$:$1$ resonance is just
starting to appear. The upper panel shows the $(u,s)$ coordinate space representations
and the lower panels show the corresponding Husimi distributions in
the $(u,p_{u})$ surface of section of the phase space.
The first two states ($E \approx 8.4$) can be assigned approximate zeroth-order quantum numbers $(n_{s},n_{u})$
whereas the last two states ($E \approx 9.0$) show perturbed nodal features. Clear difference in the phase space
nature of the eigenstates can be seen. See text for details.}
\label{FIG3}
\end{figure}

In Fig.~\ref{FIG3} we show a set of four eigenstates to illustrate an important point - as soon as
the $1$:$1$ resonance manifests itself in the phase space, the tunneling doublets start to form
and an integrable normal form approximation is insufficient to account for the splittings. 
Note that the phase space is mostly regular at the energies of interest. A much more detailed analysis
can be found in the paper by Farrelly and Uzer\cite{fu86}. 
We begin by noting\cite{note1} that the pair of eigenstates (counting from the zero-point)
$35$ and $36$ are split by about $\Delta E \approx 6.41 \times 10^{-2}$
whereas the pair $37$ and $38$ are separated by about $2.09 \times 10^{-2}$. In both cases the
splitting is smaller than the fundamental frequencies, which are of order unity. Note that the latter pair of states 
appears to be a part of the sequence in Fig.~\ref{FIG1} that starts right after the bifurcation of
the symmetric stretch periodic orbit. 
Since the original mode frequencies are nonresonant,
one can obtain a normal form approximation\cite{fu86} to the original Hamiltonian and see if the
obtained splittings can be explained satisfactorily. In other words, the observed $\Delta E$
are coming from the perturbation that couples both the modes and in such a case there is no
reason for classifying them as tunneling doublets. However,
from the coordinate space representations of the states shown in Fig.~\ref{FIG3}
one observes that there is an important difference between the two pairs of states. 
The pair $(35,36)$ seems to have a perturbed nodal structure and hence one can approximately
assign the states using the zeroth-order quantum numbers $(n_{s},n_{u})$. Inspecting the
figure leads to the assignment $(1,6)$ and $(0,7)$ for states $35$ and $36$ respectively.
If the above arguments are correct then the splitting $\Delta E_{35,36}$ should be
obtainable from the normal form Hamiltonian. Since the theory of normal forms is explained in detail
in several textbooks\cite{lichtlieb}, I will provide a brief derivation below. Begin by using the
unperturbed harmonic action-angle variables $({\bf I},{\bf \phi})$
defined via the canonical transformation (similar set for $s$ mode):
\begin{eqnarray}
u &=& \left(\frac{2I_{u}}{\omega_{u}}\right)^{1/2} \sin \phi_{u}, \\
p_{u} &=& \left(2 \omega_{u} I_{u}\right)^{1/2} \cos \phi_{u},
\end{eqnarray}
to express the original Hamiltonian as
\begin{eqnarray}
H({\bf I},{\bm \phi}) &=& \omega_{s} I_{s} + \omega_{u} I_{u} + \widetilde{\lambda} I_{s}^{1/2} I_{u}
                      \left\{\sin \phi_{s} -\frac{1}{2} \sin(\phi_{s}+2\phi_{u})
                             -\frac{1}{2} \sin(\phi_{s}-2\phi_{u}) \right\} \nonumber \\
                      &\equiv& H_{0}({\bf I}) + \widetilde{\lambda} H_{1}({\bf I},{\bm \phi}), 
\end{eqnarray}
where $\widetilde{\lambda} \equiv \lambda \sqrt{2}/(\omega_{u} \sqrt{\omega_{s}})$.
Since $H_{1}$ is purely oscillatory the angle average
\begin{equation}
\bar{H}_{1} \equiv \frac{1}{(2\pi)^{2}} \int d{\bm \phi} H_{1}({\bf I},{\bm \phi}) = 0
\end{equation}
and thus to $O(\widetilde{\lambda})$ the normal form Hamiltonian can be identified with the 
zeroth-order $H_{0}({\bf I})$ above. The first nontrivial correction arises at
$O(\widetilde{\lambda}^{2})$ and can be obtained using the $O(\widetilde{\lambda})$ generating
function
\begin{equation}
W_{1}({\bf I},{\bf \phi}) = \sum_{{\bf k} \neq {\bf 0}} \frac{i}{{\bf k} \cdot {\bf \omega}}
                             H_{1,{\bf k}}({\bf I}) e^{i {\bf k} \cdot {\bf \phi}}
\end{equation}
where $H_{1,{\bf k}}({\bf I})$ are the coefficients in the Fourier expansion of the oscillatory
part of $H_{1}({\bf I},{\bm \phi})$. One now obtains the $O(\widetilde{\lambda}^{2})$ correction
to the Hamiltonian as
\begin{equation}
K_{2} = \frac{1}{2} \overline{\{W_{1},H_{1}\}}
\end{equation}
with the bar denoting angle averaging of the Poisson bracket involving $W_{1}$ and $H_{1}$. Performing
the calculations the normal form at $O(\widetilde{\lambda}^{2})$ is obtained as
\begin{equation}
H_{N}^{(2)} = \omega_{s} I_{s} + \omega_{u} I_{u} + \widetilde{\lambda}^{2} K_{2}
\end{equation}
with
\begin{equation}
K_{2} = \frac{1}{\omega_{s} \omega_{u}^{2}} \left[\left(\frac{2 \omega_{u}}{\omega_{s}^{2}-4\omega_{u}^{2}}\right) I_{s}I_{u}-
        \left(\frac{3\omega_{s}^{2}-8\omega_{u}^{2}}{4\omega_{s}(\omega_{s}^{2}-4\omega_{u}^{2})}\right) I_{u}^{2} \right]
\end{equation}
The primitive Bohr-Sommerfeld quantization $I_{j} \rightarrow (n_{j}+1/2)\hbar$ yields the quantum eigenvalues
perturbatively to $O(\widetilde{\lambda}^{2})$. The procedure can be repeated to obtain the normal form Hamiltonian
at higher orders. For instance, Farrelly and Uzer have\cite{fu86} computed the normal form out to 
$O(\widetilde{\lambda}^{12})$ and used Pad\'{e} resummation techniques to improve in cases when the
zeroth-order frequencies are near-resonant. For our qualitative discussions, the $O(\widetilde{\lambda}^{2})$
normal form is sufficient. 

\begin{figure}
\includegraphics*[height=100mm,width=150mm]{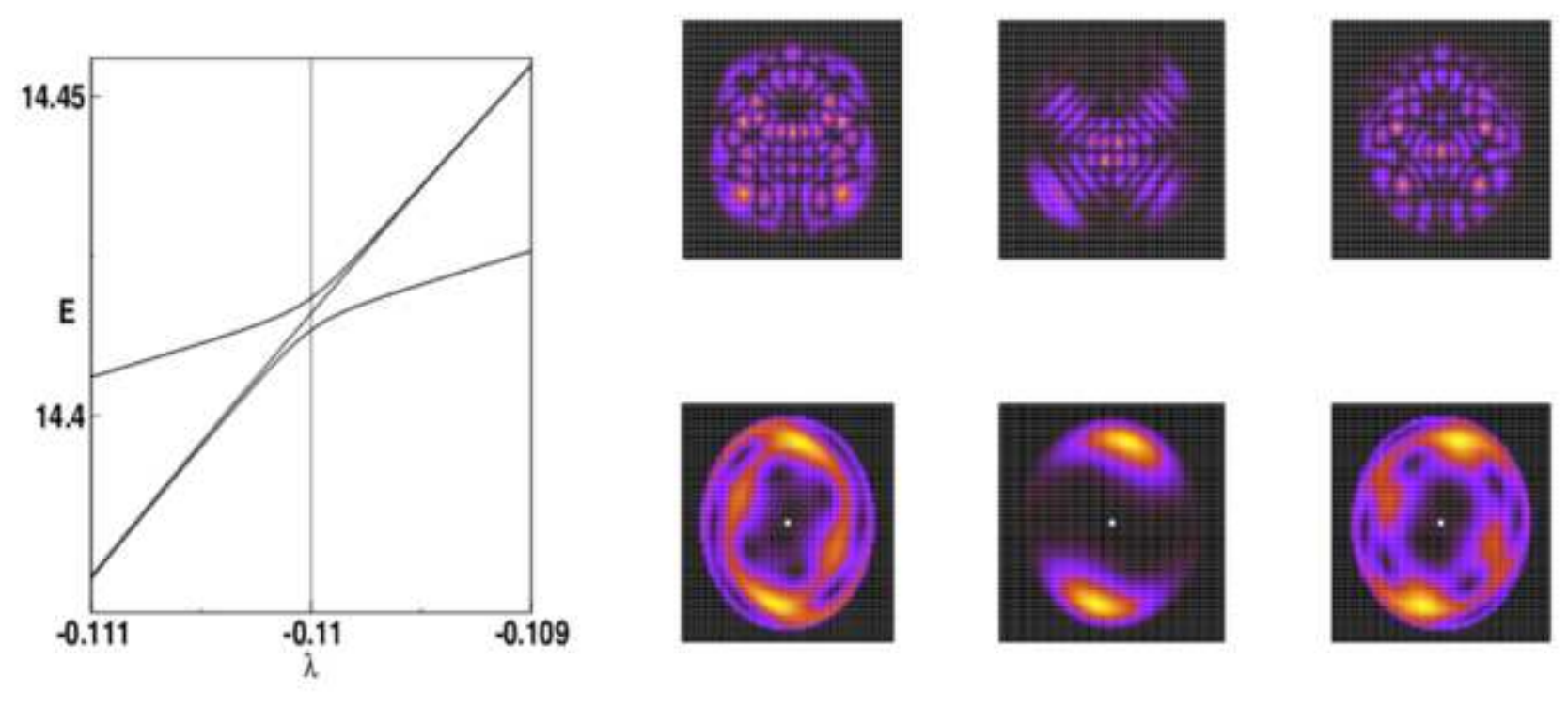}
\caption{An avoided crossing in the high energy region ($E \approx 14.4$) of the Davis-Heller system involving
three states. As in the previous figure, the $(u,s)$ and phase space Husimi representations are
also shown. The first and the last states appear to be strongly mixed states. The middle state
seems relatively cleaner. Does chaos play a role in this case? Detailed discussions in the text.}
\label{FIG4}
\end{figure}

Notice that the normal form Hamiltonian above is integrable since it is ignorable in the angle variables. Indeed, using the
normal form and the approximate assignments of the eigenstates shown in Fig.~\ref{FIG3} one finds
$\Delta E_{35,36} \approx 7.4 \times 10^{-2}$ which is in fair agreement with the actual numerical value. Thus, a large
part of the splitting $\Delta E_{35,36}$ can be explained classically. On the other hand, although the states
$37$ and $38$ seem to have an identifiable nodal structure (cf. Fig.~\ref{FIG3}), it is clear that they are significantly
perturbed. Nevertheless, persisting with an approach based on counting the nodes, states $37$ and $38$ can
be assigned as $(8,0)$ and $(7,1)$ respectively. Using the normal form one estimates $\Delta E_{37,38}
\approx 5.12 \times 10^{-2}$ and this is about a factor of two larger than the numerically computed value.
Thus, the splitting in this case is not accounted for solely by classical considerations. One might argue that
a higher order normal form might lead to better agreement, but the phase space Husimi distributions of the
eigenstates shown in Fig.~\ref{FIG3} suggests otherwise. The Husimi disributions, when compared to
the classical phase spaces shown in Fig.~\ref{FIG2}, clearly show that the pair
$(37,38)$ are localized in the newly created $1$:$1$ resonance zone. Hence, this pair of states is directly
influenced by the nonlinear resonance and the splitting between them cannot be accurately described by
the normal form Hamiltonian. Indeed, as discussed by Farrelly and Uzer\cite{fu86}, in this instance one needs to 
consider a resonant Hamiltonian which explicitly takes the $1$:$1$ resonance into account. 
This provides a clear link between dynamical tunneling, appearance of closely spaced doublets and
creation of new, in this case a nonlinear resonance, phase space structures.
The choice of states in Fig.~\ref{FIG3} is different from what is usually shown as the standard example
for dynamical tunneling pairs. However, in the discussion above the states were chosen intentionally with
the purpose of illustrating the onset of near-degeneracy due to the formation of a nonlinear resonance.
In a typical situation involving near-integrable phase spaces there are several such resonances ranging
from low orders to fairly high orders. The importance of a specific resonance depends sensitively on
the effective value of the Planck's constant\cite{Schl06,lbks10}. 
Indeed, detailed studies\cite{mes06,Kes03,Kes05} have shown that excellent
agreement with numerically computed splittings can be obtained if proper care is taken to
include the various resonances. 
Clear and striking examples in this context can be found in the contributions by
Schlagheck {\it et al.} and B\"{a}cker {\it et al.} in this volume. 

To finish the discussion of
the Davis-Heller system I show an example which involves states forming tunneling pairs that are fairly complicated
both in terms of their coordinate space representations as well as their phase space Husimi distributions. The
example involves three states $102$, $103$, and $104$ around $E \approx 14.4$, which is rather close to
the dissociation energy.
In the original work\cite{dh81} Davis and Heller noted that
the splittings seem to increase by an order of magnitude in this high energy region.
Notably, they commented that ``{\em perhaps the degree of irregularity between the regular regions plays some part}''.
In Fig.~\ref{FIG4} the variation of energy levels with the coupling parameter $\lambda$ is shown. One can
immediately see that the three states are right at the center of an avoided crossing. The coordinate space
representations of the states shows extensive mixing for states $102$ and $104$ while the state $103$ seems
to be cleaner. The Husimi distributions for the respective states conveys the same message. 
Comparing to the phase space sections shown in Fig.~\ref{FIG2} it is clear that the Husimis for state $102$
and $104$ seem to be ignoring the classical regular-chaotic divison - a clear indication of the quantum nature
of the mixing. Although it is not possible to strictly assign these states as chaotic, a closer inspection
does show substantial Husimi contribution in the border between the regular and chaotic regions. Interestingly,
the pairwise splitting between the states is nearly the same but localized linear combinations
of any two states exhibits two-level dynamics. Thus, the situation here is not of the generic\cite{btu93} chaos-assisted
tunneling one wherein one of the states is chaotic and interacts with the other two regular states. Nevertheless,
linear combinations of the three states reveal (not shown here) that they are mixed with each other. A look at
the coordinate and phase space representations of states $102$ and $104$ in Fig.~\ref{FIG4} reveals that
a different kind of state is causing the three-way interaction. It turns out that in the Davis-Heller model
there is another class of symmetry-related pairs that appear when the unsymmetric mode becomes unstable.
In the original work they were refered to as ``circulating" states which displayed much larger splitting then
the so-called local mode pairs. The case shown in Fig.~\ref{FIG4} involves one of the circulating pairs
interacting with the usual local mode doublet leading to the complicated three-way interaction. 
The subtle nature of this interaction is evident from the coordinate space representation of state $103$
which exhibits a broken symmetry. Interestingly, and as far as I can tell,
there has been very little understanding of such three-state interactions in the Davis-Heller system
and further studies are needed to
shed some light on the phase space nature of the 
relevant eigenstates.

\section{Dynamical tunneling and control: two examples}

In the previous section the intimate connection between dynamical tunneling and phase space
structures was introduced. The importance of a nonlinear resonance and a hint of the role played by
the chaos (cf. Fig.~\ref{FIG4}) is evident. Several other contributions in this volume discuss the importance
of various phase space strucures using different models, both continuous Hamiltonians and discrete maps.
In the molecular context, various mode-mode resonances play a critical role in the process of
intramolecular vibrational energy redistribution (IVR)\cite{ivr1}. The phenomenon of IVR is at the heart of
chemical reaction dynamics and it is now well established that molecules at high levels of
excitation display all the richness, complexity and subtelty that is expected from nonlinear
dynamics of multidimensional systems\cite{ivr2}. In this context, 
dynamical tunneling is an important agent\cite{Kes07,Hel95}
of IVR and state mixing for a certain class of initial states (akin to the so called NOON states\cite{noon})
which are typically prepared by
the experiments. The importance of the anharmonic resonances to IVR and the regimes wherein
dynamical tunneling, mediated by these resonances, is expected to be crucial is described
in some detail in this volume by Leitner. The phase space perspective on Leitner's viewpoint
can be found in a recent review\cite{Kes07} (see also Heller's contribution in the present volume). In this regard
it is interesting to note that there has been a renaissance of sorts in chemical dynamics
with researchers critically examining the validity of the two pillars of
reaction rate theory\cite{lh09} - transition state theory (TST) and the
Rice-Ramsperger-Kassel-Marcus (RRKM) theory. Since both theories
have classical dynamics at their foundation, advances in our
understanding of nonlinear dynamics and continuing efforts to
characterize the phase space structure of systems with three or more
degrees of freedom are beginning to yield crucial mechanistic
insights into the dynamics\cite{nhim1,nhim2}. At the same time, rapid advances in experimental techniques
and theoretical understanding of the reaction mechanisms has led researchers to
focus on the issue of controlling the dynamics of molecules. What implications might
dynamical tunneling have on our efforts to control the atomic and molecular dynamics?
In the rest of this article I focus on this issue and use two seemingly simple 
and well studied systems as examples
to highlight the role of dynamical tunneling in the context of coherent control.
Both examples are in the context of periodically driven systems and I refer the reader to
the work of Flatt\'{e} and Holthaus\cite{fh96} for an exposition of the close
quantum-classical correspondence in such systems.

\subsection{Driven quartic double well: chaos-assisted tunneling}

Historically, an early indication that dynamical tunneling could be sensitive to
the chaos in the underlying phase space came from the study of strongly driven
double well potential by Lin and Ballentine\cite{lb90}. The model Hamiltonian in this case
can be written down as
\begin{equation}
H(x,p;t) = H_{0}(x,p) + \lambda_{1} x \cos(\omega_{F}t),
\label{linbalham}
\end{equation}
with $\omega_{F}$ being the frequency of the monochromatic field (henceforth refered to
as the driving field) and the unperturbed part
\begin{equation}
H_{0}(x,p) = \frac{1}{2M} p^{2} + Bx^{4}-Dx^{2},
\label{lbh0ham}
\end{equation}
is the Hamiltonian corresponding to a double well potential with two symmetric minima at
$x = \pm(D/2B)^{1/2}$ and a maximum at $x=0$.
Following the original work\cite{lb90}, the parameters of the unperturbed system (assuming atomic units)
are taken to be $M=1$, $B=0.5$, and $D=10$ for which the potential has a barrier height
$V_{B}=50$ and supports about eight tunneling doublets. As is well known, in the absence of the
driving field, a wavepacket prepared in the left well can coherently tunnel into the right well
with the time scale for tunneling being inversely proportional to the tunnel splitting. This
unperturbed scenario, however, is significantly altered in the presence of a strong driving field.
In the presence of a strong field the phase space of the system exhibits large scale chaos coexisting with
two symmetry-related regular regions.
Lin and Ballentine observed that a coherent state localized in one of the regular region tunnels 
to the other symmetry-related regular region on timescales
which are orders of magnitude smaller than in the unperturbed case.
It was suspected that the extensive chaos in the system might be assisting the tunneling process.

\begin{figure}
\includegraphics*[height=140mm,width=160mm]{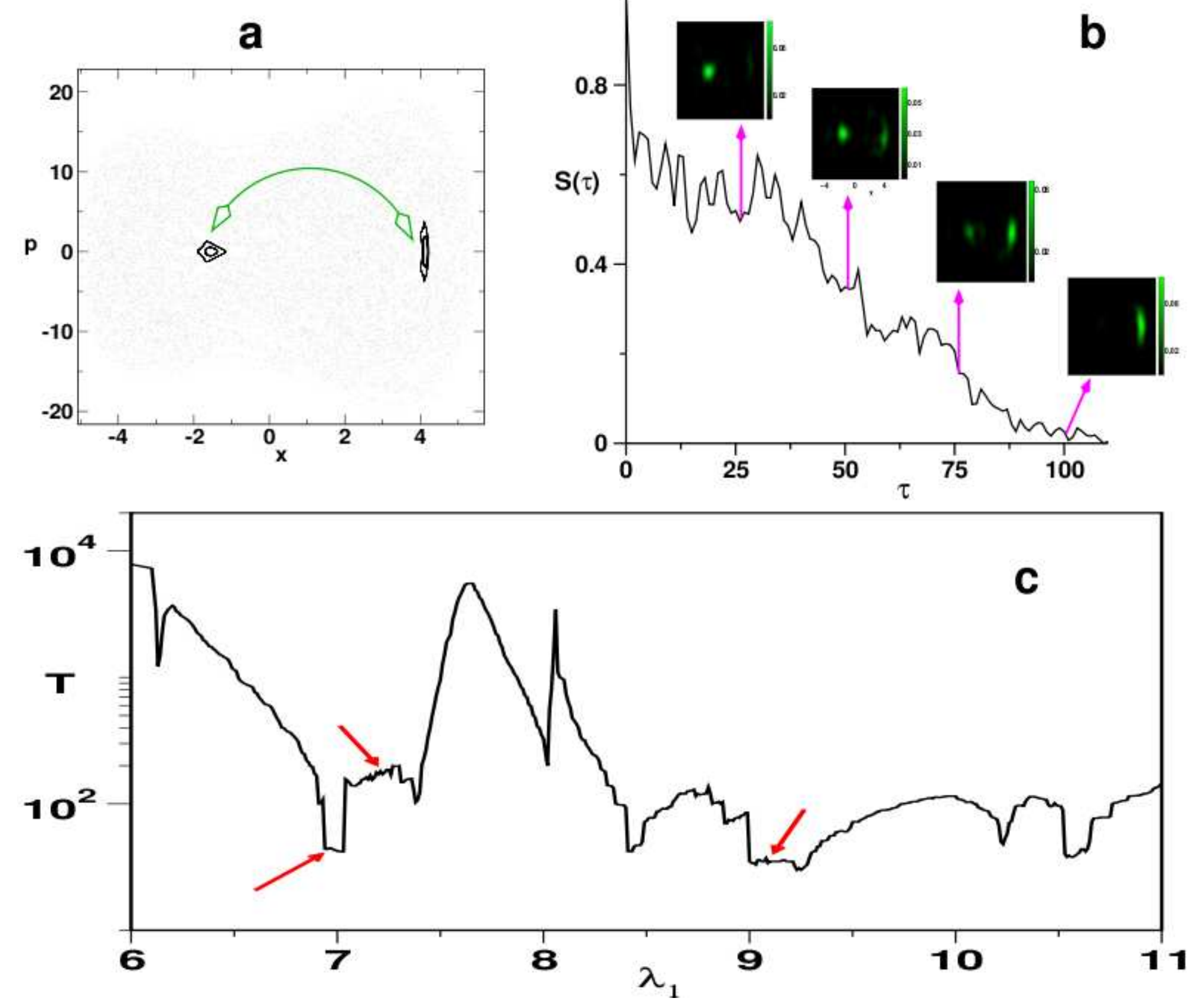}
\caption{(a) Phase space for the driven double well system with field strength $\lambda_{1}=10$. A
coherent state localized in the left regular island tunnels (indicated by a green arrow) over to
the right regular island. The chaotic regions (gray) have been suppressed for clarity. 
(b) Monitoring the survival probability of the initial state to determine
the timescale of tunneling ($\sim$ $100 \tau$ in this case). 
Snapshots of the evolving Husimi distributions are also shown
at specific intervals, with green (gray) indicating maxima of the distributions.
(c) The decay time determined as in (b) for a variety of field strengths with the initial
state localized in the left island. Note the fluctuations in the decay time over several orders
of magnitude despite very similar nature of the phase spaces over the entire range of the driving
field strength. The arrows highlight some of the plateau regions which are crucial for 
bichromatic control.}
\label{FIG5}
\end{figure}

In order to illustrate the tunneling process Fig.~\ref{FIG5}a shows the stroboscopic
surface of section for the case of strong driving with $\lambda_{1}=10$. Note the extensive chaos and
the two regular islands (left and right) in the phase space. A coherent state 
$|z\rangle \equiv |x_{0},p_{0}\rangle$ 
is placed in the center of the left island and time evolved using the Floquet approach 
which is ideally suited for time-periodic driven systems. In this instance one is interested in the time
at which the coherent state localized on the left tunnels over to the regular region on the right. In order
to obtain this information it is necessary to compute the survival probability of the
initial coherent state. Briefly, 
Floquet states \{$|\chi_{n}\rangle$\} are eigenstates of the
Hermitian operator $H-i\hbar \partial/\partial t$ 
and form a complete orthonormal basis.
An arbitrary time-evolved state $|z(t)\rangle$ can
be expressed as
\begin{equation}
|z(t)\rangle = \sum_{n}A_{n}e^{-iE_{n} t} |\chi_{n}(t)\rangle,
\end{equation}
with $E_{n}$ being the quasienergy associated with the Floquet state $|\chi_{n}\rangle$.
The expansion coefficients $A_{n}$ are independent of time
and given by
\begin{equation}
A_{n}=\langle \chi_{n}(0)|z(0)\rangle,
\end{equation}
yielding the expansion
\begin{equation}
|z(t)\rangle = \sum_{n} e^{-iE_{n}t} |\chi_{n}(t)\rangle
\langle \chi_{n}(0)|z(0)\rangle.
\end{equation}
Measuring time in units of field period ($T_{f}$) and owing to the periodicity
of the Floquet states, $|\chi_{n}(t)\rangle=|\chi_{n}(t+T_{f})\rangle$,
the above equation simplifies to
\begin{equation}
|z(\tau)\rangle = \sum_{n} e^{-iE_{n}\tau} |\chi_{n}(0)\rangle
\langle \chi_{n}(0)|z(0)\rangle \equiv \hat U(\tau)|z(0)\rangle,
\label{zper}
\end{equation}
with $\tau \equiv k T_{f}$ and integer $k$.

The time evolution operator 
\begin{equation}
\hat U(\tau)= \sum_{n} e^{-iE_{n}\tau} |\chi_{n}(0)\rangle
\langle \chi_{n}(0)|,
\end{equation}
is determined by successive application
of the one-period time evolution operator $U(T_{f},0)$ {\it i.e.},
\begin{equation}
 \hat {U}(kT_{f},0)=[\hat {U}(T_{f},0)]^k.
\end{equation}
This allows us to express the survival probability of the initial coherent state in terms of
Floquet states as
\begin{eqnarray}
S(\tau) &\equiv& |\langle z(0)|z(\tau) \rangle|^{2} \nonumber \\
&=&\left|\sum_{n} e^{-iE_{n}\tau} \langle z(0)|\chi_{n}(0)\rangle
\langle \chi_{n}(0)|z(0)\rangle \right|^{2} \nonumber \\
&=& \sum_{m,n}^{} p_{zn}p_{zm} e^{-i(E_{n}-E_{m})\tau},
\label{survprob}
\end{eqnarray}
where the overlap intensities
are denoted by $p_{zn} \equiv |\langle z(0)|\chi_{n}(0) \rangle|^{2}$.
In order to determine the ``lifetime'' ($T$) of the initial
coherent state, we monitor the time at which
$S(\tau)$ goes to zero (minimum) at the first instance.
In other words, it is the time
when the coherent state leaves its initial position {\it i.e},
the left regular island of the phase space for the first time.
In Fig.~\ref{FIG5}b the $S(\tau)$ for the initial state of interest
is shown along with 
the snapshots of Husimi distribution at specific times.
Clearly, the initial state tunnels in about $100 T_{f}$, which suggests that chaos
assists the dynamical tunneling process. However, the issue is subtle and highlighted in Fig.~\ref{FIG5}c
which shows the decay time plot for a range of driving field strengths for an
intial state localized in the left regular island. It is important to note that
the gross features of the phase space are quite similar over the entire range. However,
Fig.~\ref{FIG5}c exhibits strong fluctuations over several orders of magnitude
and this implies that a direct association of the decay time with the extent of chaos in
the phase space is not entirely correct.

\begin{center}
\begin{figure}[t]
\includegraphics*[height=90mm,width=140mm]{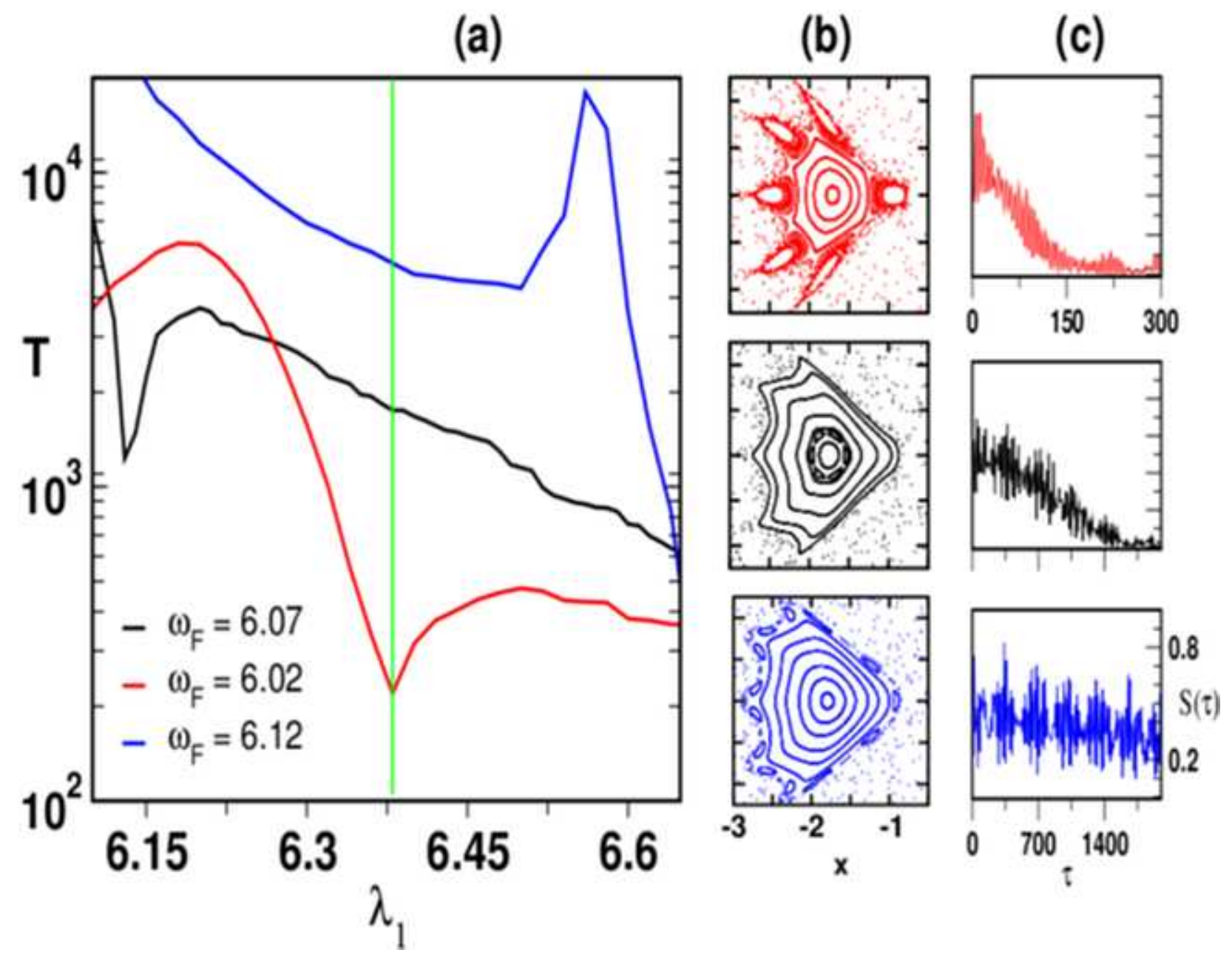}
\caption{Decay time plot for $\omega_{F}=\omega_{F}^{(0)} \equiv 6.07$
(black)
and for $\omega_{F}^{(0)}+0.05$ (blue) and $\omega_{F}^{(0)}-0.05$ (red)
as a function of $\lambda_{1}$.
The local phase structure in the vicinity of
the initial coherent state and the survival probability
for fixed $\lambda_{1}=6.4$ (indicated by green line in (a))
corresponding to red shifted,
fundamental and blue shifted field frequency
is shown in panels (b) and (c) respectively.
The $6$:$1$ field-matter resonance in case of $\omega_{F}^{(0)}-0.05$ is clearly visible
and correlates with the very short decay time as compared to the other two cases.
Note the different time axis scale in the first survival probability plot.}
\label{FIG6}
\end{figure}
\end{center}

The above discussion and results summarized in Fig.~\ref{FIG5} bring up the following key question.
What is the mechanism by which the initial state $|z\rangle$ decays out of the regular region?
In turn, this is precisely the question that modern theories of dynamical tunneling strive to answer.
According to the theory of resonance-assisted tunneling\cite{seu05,es05,Schl06}, the 
mechanism is possibily one wherein $|z\rangle$ couples to the chaotic
sea via one or several nonlinear resonances provided certain conditions are satisfied.
Specifically, the local structure of the phase space surrounding the regular region
is expected to play a critical role. The theoretical underpinnings of RAT along with several illuminating examples
can be found in the contribution by Schlagheck {\it et al.} and here we suggest a simple numerical example
which points to the importance of the local phase space structure around $|z\rangle$.

Preliminary evidence for the role
of field-matter nonlinear resonances in
controlling the decay of $|z\rangle$
is given in Fig.~\ref{FIG6}, which shows the effect of
changing the driving field frequency $\omega_{F}$ on the local phase
space structures and the decay times.
From Fig.~\ref{FIG6} 
it is apparent that detuning $\omega_{F}$ by $\pm 0.05$ leads to
a significant change in the local phase space structure and the decay time of the
initially localized state.
In Fig.~\ref{FIG6}(b), corresponding to the field frequency $\omega_{F}=6.02$,
a prominent $6$:$1$ field-matter resonance is observed. 
It is plausible that the
the decay time is only a few hundred field periods in this case
due to assistance from the nonlinear resonance.
However, the decay time increases for $\omega_{F}=6.07$
and becomes even larger by an order of magnitude
for $\omega_{F}=6.12$, due to absence of the $6$:$1$ resonance.
This indicates that decay dynamics is highly sensitive
to the changes in local phase space structure of the left regular island.
Therefore, taking into account
the relatively large order resonances, since $\hbar=1$, is unavoidable in order to
understand the decay time plot in Fig.~\ref{FIG5} and for smaller values of the
effective Planck constant one expects a more complicated behavior.
Note, however that the very high order island chain visible in the
last case in Fig.~\ref{FIG6} is unable to assist the decay. 
This is where
we believe that an extensive $\hbar$-scaling study will help in
gaining a deeper understanding of the decay mechanism. 
Such an extensive calculation can be found, for example, in
the recent work\cite{mes06} by Mouchet, Eltschka, and Schlagheck on the driven pendulum.
There is sufficient evidence\cite{mes06} in the driven pendulum system for a mechanism in which nonlinear resonances
play a central role in coupling initial states localized in regular phase space regions to
the chaotic sea. One might be able to provide a clear
qualitative and quantitative explanation for
the results in Fig.~\ref{FIG5}c based on the recent advances.

\begin{center}
\begin{figure}[t]
\includegraphics*[height=90mm,width=150mm]{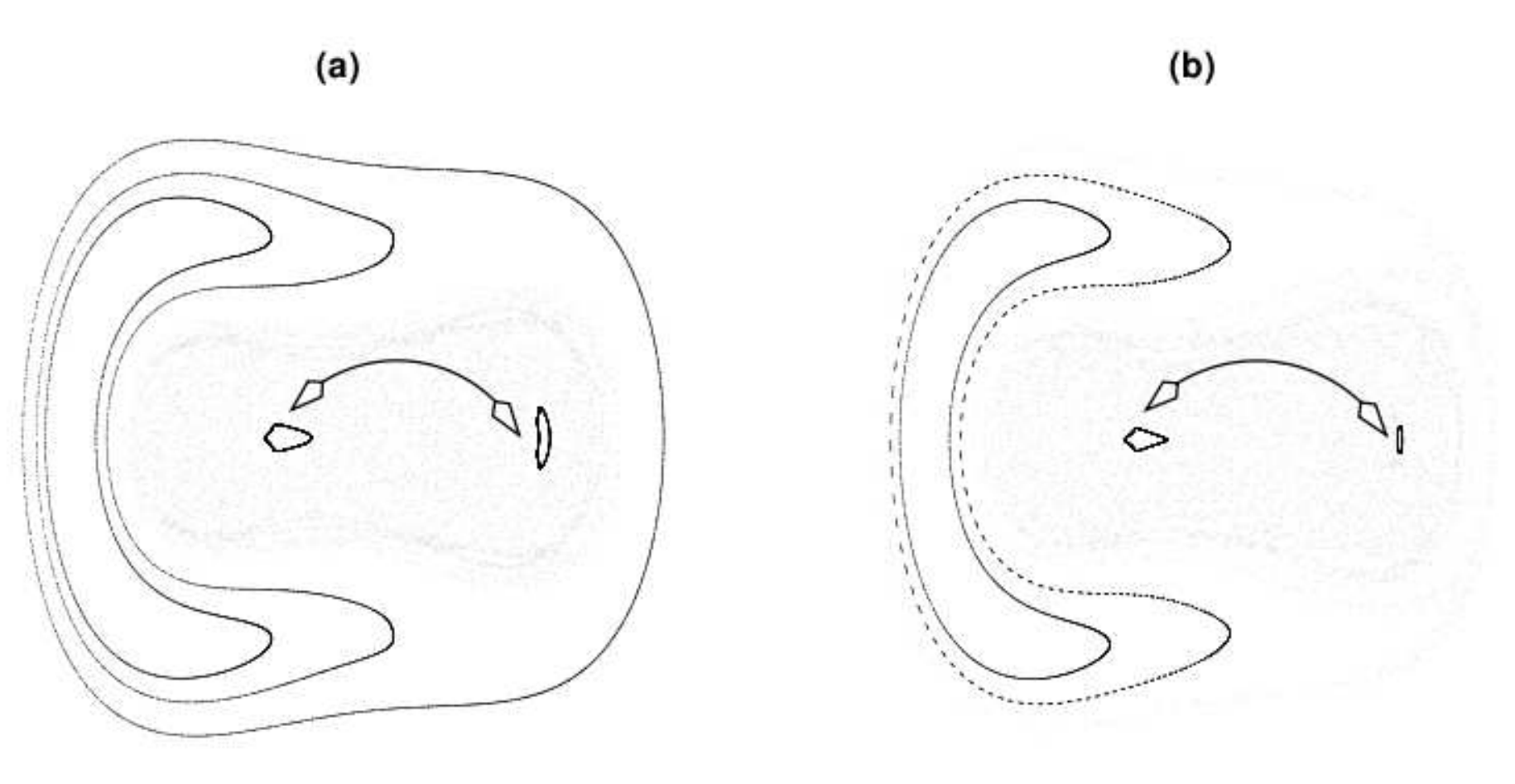}
\caption{Stroboscopic surface of sections for the bichromatically driven double well system
with the control field strength fixed at $\lambda_{2}=0.7$ and primary field strength
(a) $\lambda_{1}=9.2$ and (b) $\lambda_{1}=10.0$. As in the earlier plots, the chaotic regions 
have been suppressed (in gray) for
clarity. Note the breaking of the symmetry in both cases. However, significant suppression
of the deacy of the state localized in the left regular island happens in case (b) only.}
\label{FIG7}
\end{figure}
\end{center}

\subsubsection{CAT spoils bichromatic control}

This brings us to the second important issue - what is the role of chaos in this
dynamical tunneling process? An earlier work\cite{udh94} by Utermann, Dittrich and H\"{a}nggi on
the driven double well system showed that
there is indeed a strong correlation between the splittings of the Floquet states and the
overlaps of their Husimi distributions with the chaotic regions in the phase space. However, a different perspective
yields clear insights into the role of chaos with potential implications for coherent control.
In order to highlight this perspective I start with a simple question: {\em is it possible
to control the decay of the localized initial state with an appropriate choice of a control field?}
It is important to note that the Lin-Ballentine system parameters imply that one is dealing
with a multilevel control scenario and there has been a lot of activity over the last few years
to formulate control schemes involving multiple levels in both atomic and
molecular systems. The driven double well system has been a particular favorite in this regard,
more so in recent times due to increased focus on the 
physics of trapped Bose-Einstein condensates\cite{mo06}.
More specifically, several studies have explored the possibility of controlling
various atomic and molecular phenomenon using bichromatic fields with the relative phase between the
fields providing an additional control parameter. In the present context, for example, Sangouard {\it et al.}
exploited the physics of adiabatic passage to show that an appropriate combination of $(\omega_{F},2\omega_{F})$ field
leads to suppression of tunneling in the driven double well model\cite{prl04-sgmj-chap2}.
The choice of relative phase between the two fields
allowed them to localize the initial state in one or the other well.
However, the parameter regimes in the work by Sangouard {\it et al.} correspond to the underlying
phase space being near-integrable and hence a minimal role of the chaotic sea.

More relevant to the mixed regular-chaotic phase space case presented in Fig.~\ref{FIG5}
is an earlier work\cite{fm93} by Farrelly and Milligan wherein it was demonstrated that 
one can suppress the tunneling dynamics in a driven double well
system using a $(\omega_{F},2\omega_{F})$ bichromatic field. In other words the original Hamiltonian 
of Eq.~\ref{linbalham} is modified as following
\begin{equation}
H(x,p;t) = H_{0}(x,p) + \lambda_{1} x \cos(\omega_{F}t) + \lambda_{2} x \cos(2\omega_{F}t + \phi),
\label{farmilham}
\end{equation}
with $H_{0}(x,p)$ being the same as in Eq.~\ref{lbh0ham} and the additional $2\omega_{F}$-field is
taken to be the control field.
Moreover, modulating the turn-on time of
the control field can trap the wavepacket
in the left or right well of the double well potential.
Hence, it was argued\cite{fm93} that the tunneling dynamics in a driven double well can be controlled at will
for specific choices of the control field parameters $(\lambda_{2},\phi)$.
Note that in the presence of control field {\it i.e.,} $\lambda_{2} \ne 0$
with the relative phase $\phi=0$ the Hamiltonian in Eq.~\ref{farmilham} transforms under
symmetry operations as
\begin{eqnarray}
H\left(-x,-p;t+\frac{\pi}{\omega_{F}}\right)&=&H_{0}(x,p)-x[\lambda_{1}\cos(\omega_{F}t+\pi)+
\lambda_{2}\cos(2\omega_{F}t+2\pi)] \nonumber \\
&=&H_{0}(x,p)-x[-\lambda_{1}\cos(\omega_{F}t)+\lambda_{2}\cos(2\omega_{F}t)] \\
&\neq& H(x,p;t). \nonumber
\end{eqnarray}
Similarly, except at $\phi = \pi/2$, the discrete symmetry of the Hamiltonian
is broken under the influence of the additional $2\omega_{F}$-field.
Farrelly and Milligan thus argued\cite{fm93} that the control field with strength smaller than
the driving field will lead to localization due to the breaking of the generalized
symmetry of the Hamiltonian and the Floquet states.
The impact of a small symmetry breaking control field with $(\lambda_{2},\phi)=(0.7,0)$
can be clearly seen in Fig.~\ref{FIG7} in terms of the
changes in the classical phase space structures. However, there is a subtlety which
is not obvious upon inspecting the phase spaces shown in Fig.~\ref{FIG7} for two different but close
values of the driving field strength. In case of $\lambda_{1}=9.2$ corresponding to
Fig.~\ref{FIG7}(a), the control field is unable to suppress the
decay of the initial state $|z\rangle$ localized in the left regular region. On the other hand,
Fig.~\ref{FIG7}(b) corresponds to $\lambda_{1}=10$ and computations show that the control
field is able to suppress the decay of $|z\rangle$ to an appreciable extent. Thus, although
in both cases the control field breaks the symmetry and the resulting phase spaces show very
similar structures, the extent of control exerted by the $2\omega_{F}$-field is drastically different.

\begin{figure}[t]
\begin{center}
\includegraphics[height=100mm,width=160mm]{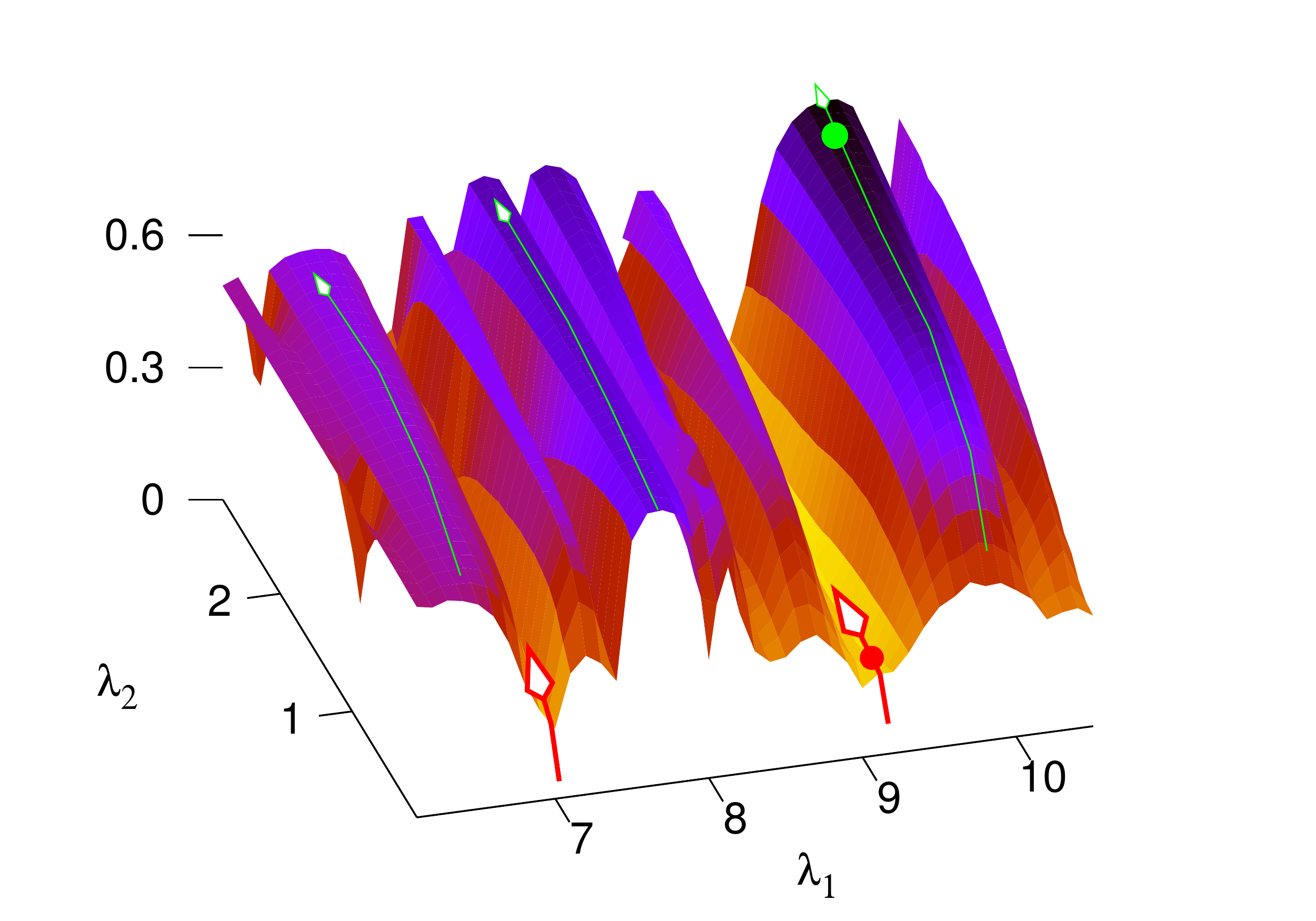}
\caption{Time-smoothed survival probability representation
of the control landscape for $\phi=0$ as a function of the 
field parameters $(\lambda_{1},\lambda_{2})$. The initial state
in every case is localized in the left regular region in the classical phase space. Notice the
convoluted form of the landscape with the regions of low
probability indicating little to no control (red thick arrows). The green lines (thin arrows) indicate regions
where a high degree of control can be achieved.}
\label{FIG8}
\end{center}
\end{figure}

The discussions above and the results summarized in Fig.~\ref{FIG7} and Fig.~\ref{FIG6} clearly indicate
that the decay of $|z\rangle$ out of the left regular region can be very different
and deserves to be understood in greater detail.
In fact, computations show that such cases of complete lack of control are present for
other values of $\lambda_{1}$ as well. This is confirmed by inspecting Fig.~\ref{FIG8}
which shows the control landscape for the bichromatically driven double well in the 
specific case of $\phi=0$. Other choices for $\phi$ also show similarly convoluted landscapes.
There are several ways of presenting a control landscape and in Fig.~\ref{FIG8}
the time-smoothed survival probability (cf. Eq.~\ref{survprob})
\begin{equation}
\langle S \rangle = \frac{1}{\tau} \int_{0}^{\tau} d \tau' S(\tau')
\end{equation}
associated with the initial state $|z\rangle$ is used to map the landscape
as a function of the field strengths $(\lambda_{1},\lambda_{2})$.
Note that the choice of $\langle S \rangle$ to represent the
landscape is made for convenience; the decay time is a better choice
which requires considerable effort but the gross qualitative features of
the control landscape do not change upon using $\langle S \rangle$.
Large (small) values of $\langle S \rangle$ indicate that the decay dynamics
is suppressed (enhanced). It is clear from Fig.~\ref{FIG8}
that the landscape comprises of regions of control interspersed
with regions exhibiting lack of control. Such highly convoluted
features of control landscape are a consequence of the simple bichromatic choice for
the control field and the nonlinear nature of the corresponding classical dynamics.
From a control point of view, there are regions on the
landscape for which a monotonic increase of $\lambda_{2}$ leads to
increasing control. Interestingly, the lone example of control illustrated in Farrelly and Milligan's
paper\cite{fm93} happens to be located on one of the prominent hills on the control landscape (shown as 
a green dot in Fig.~\ref{FIG8}).
A striking feature that can be seen in Fig.~\ref{FIG8} is the deep valley around $\lambda_{1}=9.2$
which signals an almost complete lack of control even for significantly large strengths of
the $2\omega_{F}$-field. The valleys in Fig.~\ref{FIG8} correspond precisely to the plateaus seen in
the decay time plot shown in Fig.~\ref{FIG6} for $\lambda_{2}=0$ (red arrows).
It is crucial to note that this ``wall of no control" is robust
even upon varying the relative phase $\phi$ between the driving and the control field. 

\begin{figure}[t]
\begin{center}
\includegraphics[height=120mm,width=140mm]{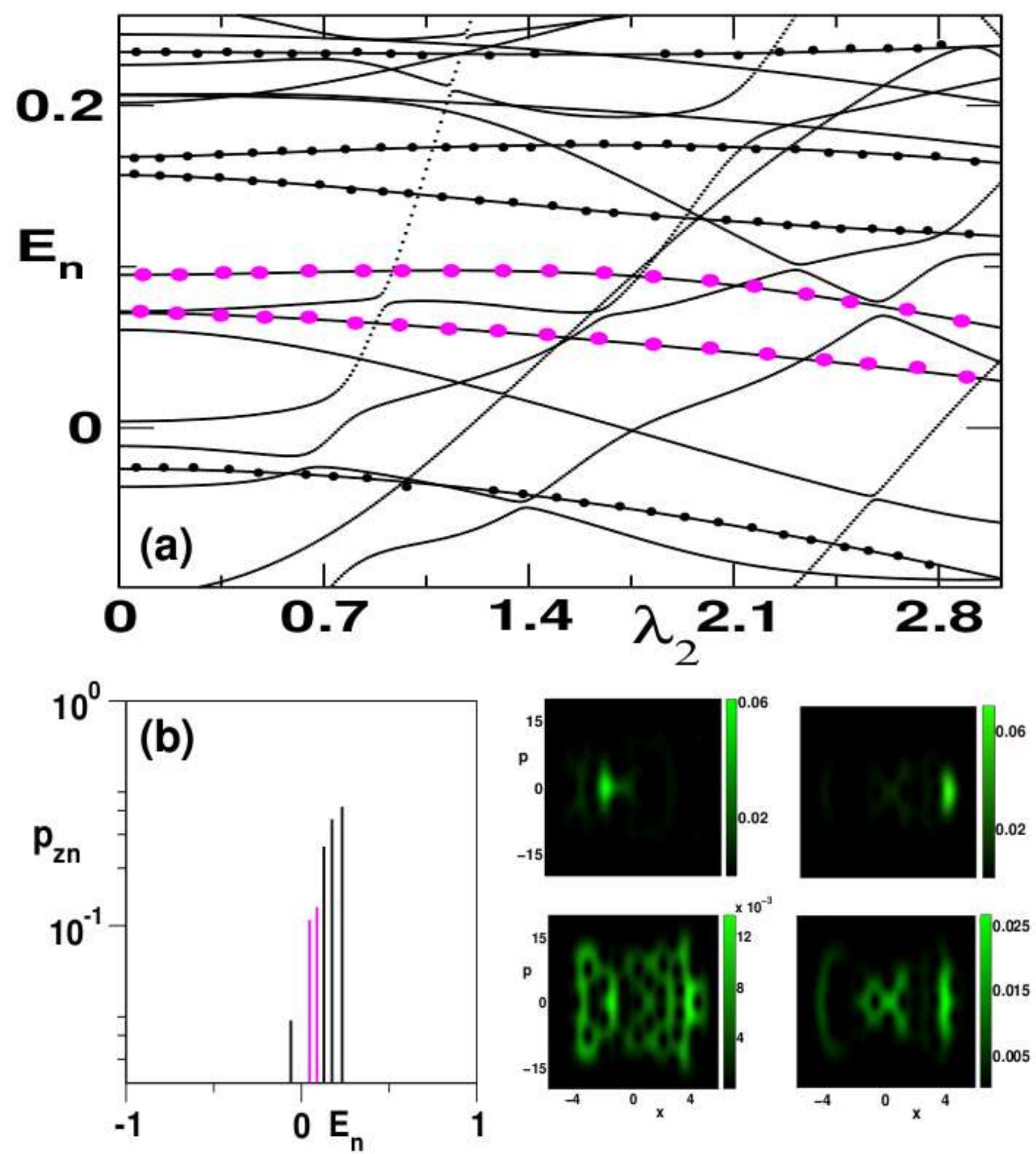}
\caption{(a) Variation of the Floquet quasienergies with $\lambda_{2}$. The primary
driving field strength is fixed at
$\lambda_{1} = 9.2$. Six states that have appreciable
overlap with $|z\rangle$ are highlighted by circles.
(b) Overlap intensity $p_{zn}$ for $\lambda_{1}=9.2$ and
$\lambda_{2}=2.1$ indicates multilevel interactions involving
the states shown in (a). Husimi distribution function of the
Floquet states regulating the decay of $|z\rangle$ are also shown.
Notice that the nature of delocalized states (magenta)
does not change much with $\lambda_{2}$.}
\label{FIG9}
\end{center}
\end{figure}

Insights into the lack of control for driving field strength $\lambda_{1}=9.2$ (and other
values as well which are not discussed here)
can be obtained by studying the
variation of the Floquet quasienergies
with $\lambda_{2}$, the control field strength.
In Fig.~\ref{FIG9} the results of such a computation are shown
and it is immediately clear that even in the presence of $2\omega_{F}$-control field
six states contribute to the decay of $|z\rangle$ over the
entire range of $\lambda_{2}$.
This is further confirmed in Fig.~\ref{FIG9}(b)
where the plot of the overlap intensities shows multiple Floquet states participating
nearly equally in the decay dynamics of the initial state.
The final clue comes from inspecting the Husimi distributions 
shown in Fig.~\ref{FIG9}, highlighting the phase space delocalized nature of
some of the participating states. Despite the symmetry of the tunneling doublets being broken due to the
bichromatic field, two or more of the participating states are extensively delocalized in
the chaotic reigons of the phase space. Moreover, 
the participation of the chaotic Floquet states persists even
for larger values of $\lambda_{2}$.
Hence, using the symmetry breaking property of the
$2\omega_{F}$-field for control purposes is not
very effective when chaotic states are participating in the dynamics.
{\em Therefore, the lack of control,
signaled by plateaus in Fig.~\ref{FIG6} and the valleys in Fig.~\ref{FIG8},
is due to the dominant participation by
chaotic states {\it i.e.,} chaos-assisted tunneling.}
The plateaus arise due to the fact that the coupling between the
localized states and the delocalized states
vary very little with increasing control field strength $\lambda_{2}$ -
something that is evident from the Floquet level motions shown in
Fig.~\ref{FIG9} and established earlier by Tomsovic and Ullmo in their
seminal work on chaos-assisted tunneling
in coupled quartic oscillators\cite{tu94}.
It is important to note that for $\lambda_{2}=0$
the chaotic states, as opposed to the regular states,
do not have a definite parity. Consequently, the presence of the
$2\omega_{F}$-field does not have
a major influence on the chaotic states.
Thus, if one or more chaotic states are already influencing the dynamics
of $|z\rangle$ at $\lambda_{2}=0$ then the bichromatic control is
expected to be difficult. An earlier study\cite{lgw94} by Latka {\it et al.} on the
bichromatically driven pendulum system also suggested that the ability to control
the dynamics is strongly linked to the existence of chaotic states. 

The model problem in this section and the results point to a direct role
of chaos-assisted pathways in the failure of an attempt to bichromatically control
the dynamics. However, it is not yet clear if control strategies involving more
general fields would exhibit similar characteristics. There is some evidence in the literature
which indicates that quantum optimal control landscapes might be highly convoluted if the
underlying classical phase space exhibits large scale chaos\cite{sr91}. Nevertheless, further studies need to be done
and the resulting insights are expected to be crucial in any effort to control
the dynamics of multilevel systems. 

\subsection{Driven Morse oscillator: resonance-assisted tunneling}

The driven Morse oscillator system has served as a paradigm model for understanding the
dissociation dynamics of diatomic molecules. Studies spanning nearly three decades have explored
the physics of this system in exquisite detail. Consequently, a great deal is known about the
mechanism of dissociation both from the quantum and classical perspectives. Indeed, the
focus of researchers nowadays is to control, either suppress or enhance, the dissociation dynamics
and various suggestions have been put forward. In addition, one hopes that the ability to control
a single vibrational mode dynamics can lead to a better understanding of the complications
that arise in the case of polyatomic molecular systems wherein several vibrational modes
are coupled at the energies of interest.

Several important insights have originated from classical-quantum correspondence studies
which have established that molecular dissociation, in analogy to multiphoton
ionization of atoms,
occurs due to the system gaining energy by diffusing through the
chaotic regions of the phase space. For example, an important experimental study by
Dietrich and Corkum has shown\cite{jcp92-dc-chap3}, amongst other things, the validity of the chaotic
dissociation mechanism.
Thus, the formation of the chaotic regions due to
the overlap\cite{pr79-ch-chap3} of nonlinear resonances (field-matter),
hierarchical structures\cite{prl84-mmp-chap3} near the
regular-chaotic borders acting as partial barriers,
and their effects on quantum transport\cite{prl88-rp-chap3}
have been studied in a series of elegant papers\cite{bw861,pra87-gy-chap3,pra91-gh-chap3}.
In the context of this present article an interesting question is as follows. 
Since a detailed mechanistic understanding of the role of various phase space structures
of the driven Morse system is known, is it possible
to design local phase space barriers to effect control over the dissociation dynamics?
In particular, the central question here is whether the local phase space
barriers are also able to suppress the quantum dissociation dynamics. There is an obvious
connection between the above question and the theme of this volume - quantum mechanics
can ``shortcircuit" the classical phase space barriers due to the phenomenon of dynamical tunneling.
Thus, such phase space barriers might be very effective in controlling the classical dissociation
dynamics but might fail completely when it comes to controlling the quantum dissociation dynamics.
There is a catch here, however, since there is also the possibility that the cantori barrier in the
classical phase space may be even more restrictive in the quantum case. Thus, creation
or existence of a phase space barrier invariably leads to subtle competition
between classical transport and quantum dynamical tunneling through the barrier.
As expected, the delicate balance between the classical and quantum mechanism is determined by
the effective Planck constant of the system of interest.
An earlier detailed review\cite{grr86}
by Radons, Geisel and Rubner is highly reccomended for a nice introduction to the subject of
classical-quantum correspondence perspective on phase space transport through 
Kolmogoroff-Arnol'd-Moser (KAM) and cantori barriers. In the driven Morse oscillator case,
Brown and Wyatt showed\cite{bw861} that the cantori barriers do leave their imprint on the quantum
dissociation dynamics and act as even stronger barriers as compared to the classical system.
Maitra and Heller in their study\cite{pre00-mh-chap3} on transport through cantori in the whisker map
have clearly highlighted the classical versus quantum competition.

\begin{center}
\begin{figure}[t]
\includegraphics[height=110mm,width=140mm]{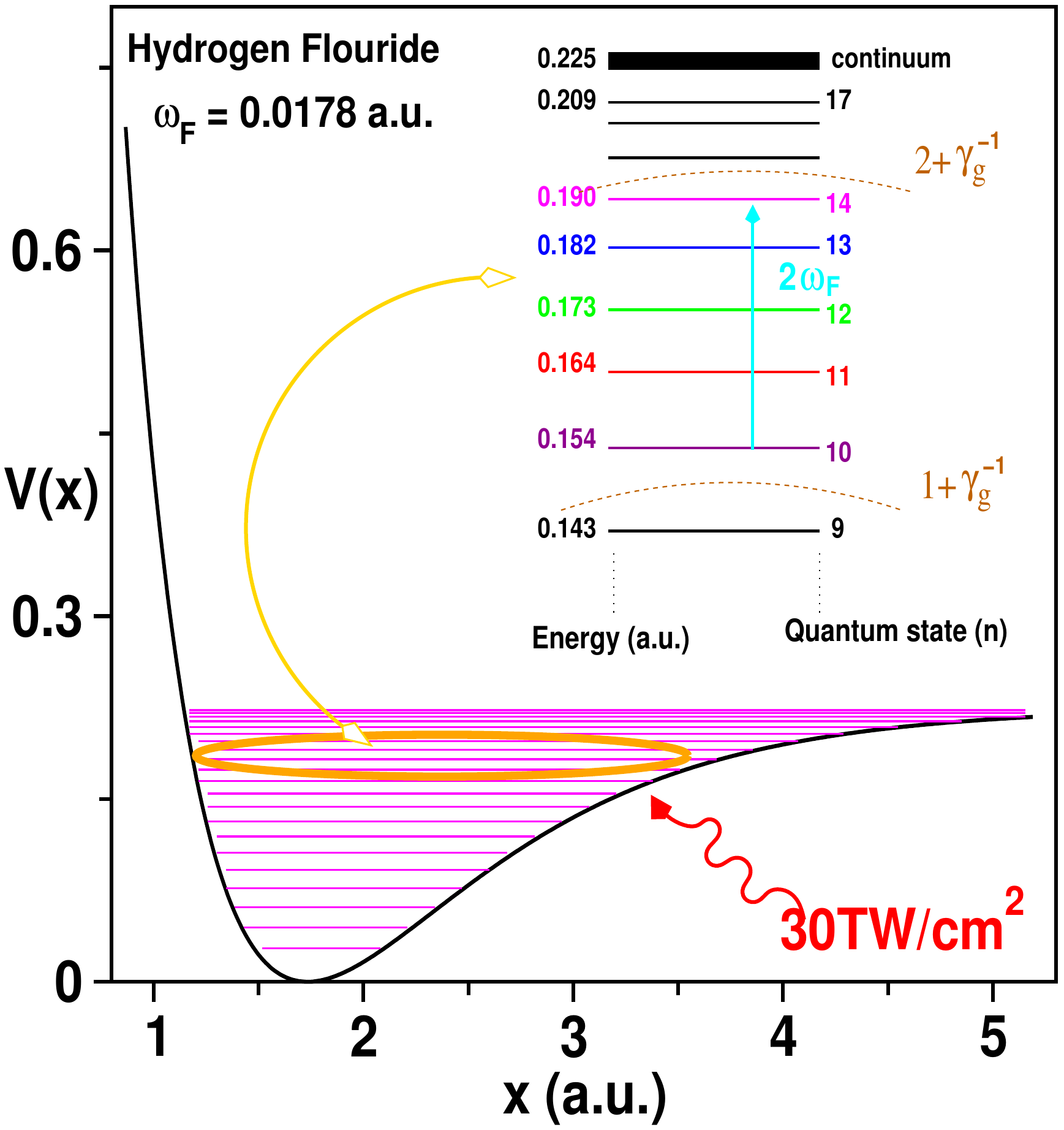}
\caption{A plot of Morse potential along with the
bound states supported by the potential well. A schematic view
of energy level diagram for Morse eigenstates under study in this
chapter is shown as inset. The laser field of
intensity $\lambda_{1}=0.0287$ and frequency
$\omega_{F}=0.0178$ connects state $n=10$ and
$n=14$ via a two-photon resonance transition. The location
of cantori in presence of the field with $\omega_{F}=0.0178$
are indicated as dotted curves in the figure.}
\label{FIG10}
\end{figure}
\end{center}

From the above discussion it is apparent that any approach to control the dissociation 
dynamics by recreating local phase space barriers will face the subtle classical-quantum
competition. In fact, it is tempting to think that every quantum control algorithm
works by creating local phase space dynamical barriers and the efficiency of the control
is decided by the classical-quantum competition. However, at this point of time there is
very little work towards making such a connection and the above statement is, at best, a
conjecture. For the purpose of this article, I turn to the driven Morse oscillator
system to provide an example for the importance of resonance-assisted tunneling in 
controlling the dissociation dynamics. 

The model system is inspired from the early
work\cite{bw861} by Wyatt and Brown (see also the work by Breuer and Holthaus\cite{bh93})
and the Hamiltonian can be written as
\begin{equation}
H(x,p;t)=H_{0}(x,p)-\mu(x)\lambda_{1} \cos(\omega_{F} t),
\label{drivmorseham}
\end{equation}
in the dipole approximation. The zeroth-order Hamiltonian
\begin{equation}
H_{0}(x,p)=\frac{1}{2M}p^{2}+D_{0}{[1-e^{{-\alpha (x-x_{e})}}]}^2,
\label{morsepot}
\end{equation}
represents the Morse oscillator modeling the anharmonic vibrations of a diatomic molecule.
In the above, $\mu(x)$ is the dipole moment function,
$\lambda_{1}$ is the strength of the laser field,
$\omega_{F}$ is the driving field frequency, and 
\begin{equation}
M=\frac{m_{1}m_{2}}{m_{1}+m_{2}},
\end{equation}
is the reduced mass of the diatomic molecule with $m_{1}$ and $m_{2}$ being the two atomic masses.
In Eq.~\ref{morsepot}, $D_{0}$ is the dissociation energy,
$\alpha$ is the range of the potential and
$x_{e}$ is the equilibrium bond length of the molecule.

Rather than attempting to provide a general account as to how RAT might interfere with
the process of control, I feel that it is best to illustrate with a realistic molecular example.
Hopefully, the generality of the arguments will become apparent later on. For the present purpose
I choose the diatomic molecule hydrogen fluoride (HF) as the specific example. Any diatomic
molecule could have been chosen but HF is studied here due to the fact that Brown and Wyatt
have already discussed the role of cantori barriers to the dissociation dynamics in some detail.
The Morse oscillator parameters for hydrogen fluoride are $D_{0}=0.225, \alpha=1.174,
x_{e}=1.7329$ and $M=1744.59$.
These parameters correspond to ground electronic state
of the HF molecule supporting $N_{B}=24$ bound states.
Note that atomic units are used for
both the molecular and field parameters with time being
measured in units of the field period $\tau_{F}=2\pi/\omega_{F}$.
The only difference between the present work and that of Brown and Wyatt has to do with the
field-matter coupling. Brown and Wyatt use the dipole function
\begin{equation}
\mu(x)=A x e^{-B{x^4}},
\label{dipo}
\end{equation}
with $A=0.4541$ and $B=0.0064$, obtained from {\it ab-initio} data on HF. Here a linear
approximation for $\mu(x)$ 
\begin{eqnarray}
\mu(x) &\approx& \mu(x_{e})+
 \left(\frac{\partial \mu}{\partial x}\right)_{x_{e}}(x-x_{e}) \nonumber \\
&\equiv& \mu(x_{e})+d_{1}(x-x_{e}),
\label{dipo1}
\end{eqnarray}
is employed with $d_{1} \approx 0.33$ in case of HF. There are quantitative differences
in the dissociation probabilities due to the linearization approximation
but the main qualitative features remain intact despite the linearization approximation. 

In Fig.~\ref{FIG10} the Morse potential for HF is shown along with a summary
of the key features such as the energy region of interest, the quantum state
whose dissociation is to be controlled, and the classical phase space
structures that might play an important role in the dissociation dynamics.
The driving field parameters are chosen as $(\omega_{F},\lambda_{1})=(0.0178,0.0287)$,
same as in the earlier work\cite{bw861}, and the field strength 
corresponds to about $30$ TW/cm$^{2} \equiv 30 \times 10^{12}$ W/cm$^{2}$.
As shown in Fig.~\ref{FIG10} (inset), the focus is on understanding and controlling the
dissociation dynamics of the $n=10$ excited Morse oscillator eigenstate. There are several
reasons for such a choice and I mention two of the most important reasons. Firstly,
the earlier study\cite{bw861} has established that $n=10$ of HF happens to be in an energy regime wherein
two specific cantori barriers in the classical phase space affect the dissociation dynamics.
Moreover, for driving frequency $\omega_{F}=0.0178$, 
two of the zeroth-order eigenstates $n=10$ and $n=14$ have unperturbed energies such that
$E_{14}-E_{10} \approx 2\hbar \omega_{F}$ and
hence corresponds to a two-photon resonant situation. Secondly, for 
a field strength of $30$ TW/cm$^{2}$, the dissociation
probability of the ground vibrational state is negligible.
Far stronger field strengths are required to dissociate the $n=0$ state and ionization
process starts to compete with the dissociation at such high intensities. Thus, in order to
to illustrate the role of phase space barriers in the
dissociation dynamics without such additional complications, the specific initial state $n=10$ is chosen. 
Incidentally, such a scenario is quite feasible since
a suitably chirped laser field can populate the $n=10$ state very efficiently
from the initial ground state $n=0$
and one imagines coming in with the monochromatic laser to dissociate the molecule.

\subsubsection{Nature of the classical phase space}

The monochromatically driven Morse system studied here has a dimensionality such that 
one can conveniently visualize the phase space in the original cartesian $(x,p)$ variables.
However, since the focus is on suppressing dissociation by creating
robust KAM tori in the phase space, action-angle
variables $(J,\theta)$ which are canonically conjugate to $(x,p)$
are convenient and a natural representation to work with.
The action-angle variables $(J,\theta)$ of the unperturbed
Morse oscillator, appropriate for the bound regions, are given by\cite{rm71}
\begin{subequations}
\begin{eqnarray}
J &=& \sqrt{\frac{2MD_0}{\alpha^2}}\left(1-\sqrt{1-E}\right) \\
\theta &=& -sgn(p) \cos^{-1}\left[\frac{1-E}{\sqrt{E}}e^{\alpha(x-x_e)}-
\frac{1}{\sqrt{E}}\right].
\end{eqnarray}
\label{actionangle}
\end{subequations}
In the above equations, $E=H_{0}/D_{0}<1$ denotes the dimensionless bound state energy,
and $sgn(p)=1$ for $p \geq 0$, $sgn(p)=-1$ for $p < 0$.
In terms of the action-angle variables it is possible to
express the cartesian $(x,p)$ as follows
\begin{subequations}\label{xpactang_chap3}
\begin{eqnarray}
x &=& x_e+\frac{1}{\alpha}\ln\left[\frac{1+\sqrt {E_{0}(J)}\cos\theta}{(1-E_{0}(J))}\right]\\
p&=&\frac{-(2MD_0)^{1/2} [E_{0}(J)(1-E_{0}(J))]^{1/2}\sin \theta}{1+\sqrt{E_{0}(J)}\cos \theta},
\end{eqnarray}
\end{subequations}
where $E_{0}(J)=H_{0}(J)/D_{0}$.
Substituting for $(x,p)$ in terms of $(J,\theta)$,
the unperturbed Morse oscillator Hamiltonian in Eq.~\ref{morsepot} is transformed into
\begin{equation}
H_{0}(J) = \omega_0\left(J-\frac{\omega_0}{4D_0} J^2\right), 
\label{hact0}
\end{equation}
where $\omega_0=\sqrt{2\alpha^2 D_0/M}$
is the harmonic frequency at the minimum.
The zeroth-order nonlinear frequency is easily obtained as
\begin{equation}
\Omega_{0}(J) \equiv \frac{\partial H_{0}}{\partial J}=\omega_{0}\left(1-\frac{\omega_{0}}{2D_{0}} J \right),
\label{nonlinfreq}
\end{equation}
and with increasing excitation {\it i.e.,} increasing action (quanta) $J$,
the non-linear frequency $\Omega_{0}(J)$ decreases monotonically and eventually vanishes,
signaling the onset of unbound dynamics leading to dissociation.

\begin{center}
\begin{figure}[t]
\includegraphics[height=110mm,width=160mm]{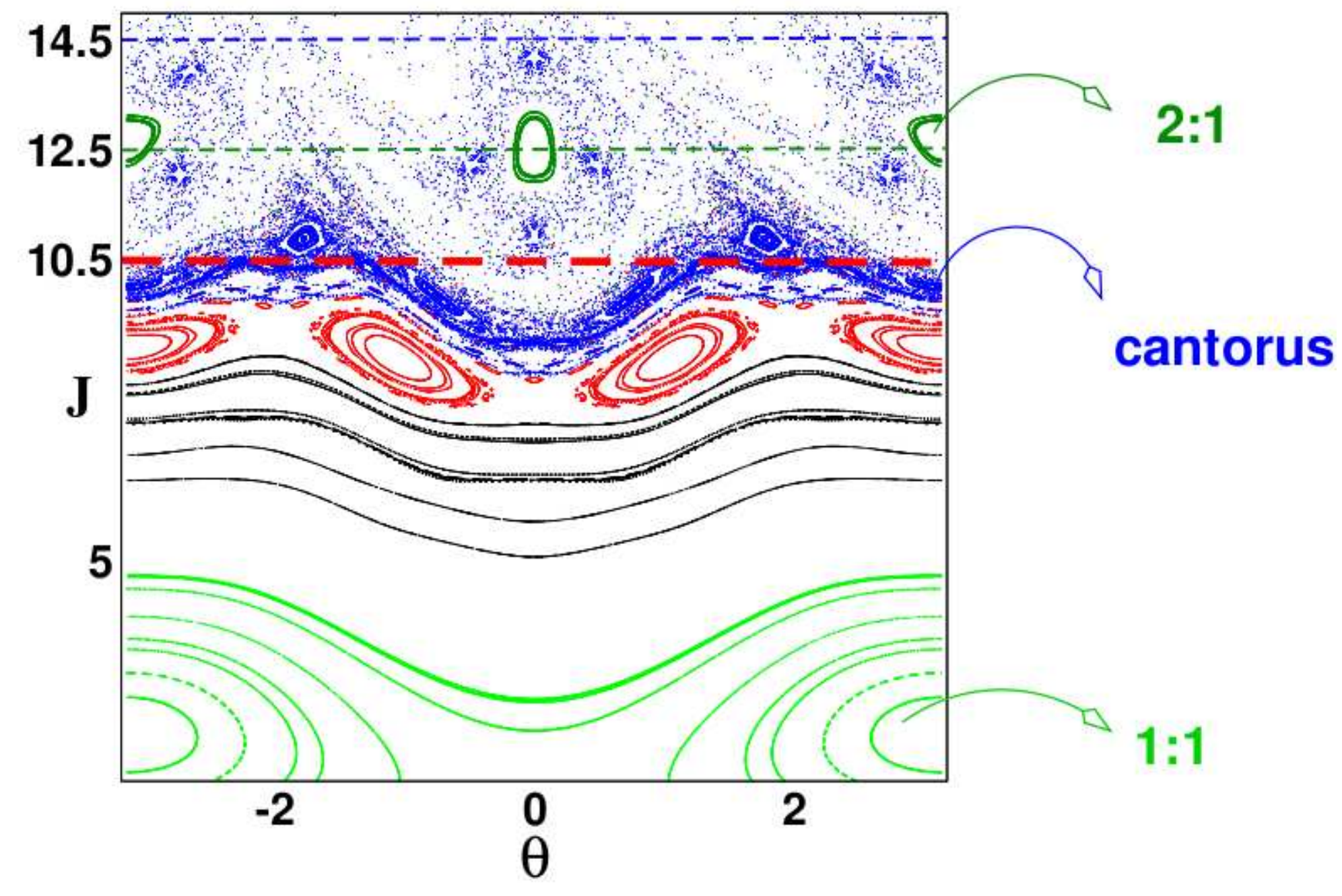}
\caption{The phase space, as a stroboscopic surface of section, for the
driven Morse system with laser field of
intensity $\lambda_{1}=0.0287$ and frequency
$\omega_{F}=0.0178$ corresponding to a on-resonance situation.
The $2$-photon resonance is clearly seen in the phase space and marked as
$2$:$1$ resonance in dark green. The initial Morse eigenstate $n=10$ (thick
dashed line)
is situated rather close to a cantorus (indicated), $\omega_{F}/\Omega_{0}(J) = 1+\gamma_{g}^{-1}$
with $\gamma_{g} \equiv (1+\sqrt{5})/2$ being the golden ratio.
The $n=10$ state is connected to the $n=14$ Morse eigenstate via the $2$:$1$ resonance. 
Note that the states $n=10,12$, and $14$ are symmetrically located about the resonance
with $n=12$ being localized in the resonance.}
\label{FIG11}
\end{figure}
\end{center}

The driven system can now be expressed in terms of the variables $(J,\theta)$ as
\begin{equation}
H(J,\theta;t) = H_0(J)-\frac{\epsilon}{\alpha}\ln\left[\frac{1+\sqrt {E_{0}(J)}\cos\theta}
{(1-E_{0}(J))}\right]\cos(\omega_{F}t).
\label{prelim_hamact}
\end{equation}
In addition, since $x$ is a periodic function of $\theta$, one
has the Fourier expansion
\begin{equation}
x = 2 \left[V_0(J)
+\sum_{n=1}^{\infty}V_n(J)\cos(n\theta) \right].
\end{equation}
As a consequence the driven Hamiltonian in Eq.~\ref{prelim_hamact}
can be written as
\begin{equation}
H(J,\theta;t) = H_0(J)-\epsilon v(J,\theta;t),
\label{hamact2}
\end{equation}
where the matter-field interaction term is denoted as
\begin{equation}
v(J,\theta;t) = 2 \left[V_0(J)
+\sum_{n=1}^{\infty}V_n(J)\cos(n\theta) \right]\cos(\omega_F t), \label{pertact} \nonumber
\end{equation}
The Fourier coefficients $V_{0}(J)$ and $V_{n}(J)$
are known analytically and given by
\begin{subequations}\label{potfour}
\begin{eqnarray}
V_{0}(J)&=&\frac{1}{2\alpha}\ln\left[\frac{D_{0}+\sqrt{{D_{0}}^2-D_{0}E_{0}(J)}}
{2(D_{0}-E_{0}(J))}\right], \label{eqn_four1}\\
V_{n}(J)&=&\frac{(-1)^{n+1}}{\alpha n}{\left[\frac{\sqrt{D_{0}E_{0}(J)}}{D_{0}+
\sqrt{D_{0}^{2}-D_{0}E_{0}(J)}}\right]}^{n} \label{eqn_four2}.
\end{eqnarray}
\end{subequations}

The stroboscopic surface of section in the $(J,\theta)$ variables is shown in
Fig.~\ref{FIG11} and is a typical mixed regular-chaotic phase space.
A few important points are worth noting at this stage. Firstly,
the initial state of interest is located close to a cantorus with
$\omega_{F}:\Omega_{0}(J) = 1+\gamma_{g}^{-1}$ with $\gamma_{g} \equiv (1+\sqrt{5})/2$ being the golden ratio.
The importance of this cantorus to the resulting dissociation dynamics of the $n=10$ state
was the central focus of the work by Brown and Wyatt\cite{bw861}. In particular,
the extensive stickiness around this region can be clearly seen and hence one expects 
nontrivial influence on the classical dissociation dynamics as well. Secondly, a prominent
$\omega_{F}$:$\Omega_{0}(J) = 2$:$1$ nonlinear resonance is also observed in the phase space
and represents the classical analog of the quantum $2$-photon resonance condition. Interestingly,
the area of this resonance is about $\hbar$ and, therefore, can support one quantum state.
It turns out that the Husimi density of the Morse state $n=12$ is localized inside the
$2$:$1$ resonance island. In addition, the states $n=10$ and $n=14$ are nearly symmetrically located
about the $2$:$1$ resonance. 
A way to see this is to use secular pertubation
theory on the driven Morse Hamiltonian. One can show that in the vicinity of the
$2$:$1$ resonance (cf. Fig.~\ref{FIG11}) an effective pendulum Hamiltonian
\begin{equation}
H_{eff}(J,\phi) \simeq \frac{1}{2\tilde{m}_{2:1}}{(\Delta J)}^2 +
2\tilde{V}_{2:1}(J_{2:1}) \cos(2\phi),
\label{pendham}
\end{equation}
is obtained with $\Delta J = J-J_{2:1}$,
$\tilde{m}_{2:1}=2D_{0}/\omega_{0}^{2}$ and
$V_{2:1}(J)=\epsilon V_{1}(J)/2$. The resonant action
\begin{equation}
J_{2:1}=\frac{2 D_{0}}{\omega_{0}}\left[1-\frac{\omega_{F}}{2 \omega_{0}}\right] \approx 12.6,
\end{equation}
for the parameters used in this work. Using action values $J=10.5$ (quantum state $n=10$) and $J'=14.5$ (quantum
state $n'=14$) the energy difference is calculated as
\begin{eqnarray}
|E_{J}-E_{J'}|&=&\left|\frac{1}{2\tilde{m}_{2:1}} (J-J')(J+J'-2J_{2:1}) \right| \\
               &\approx& 3.2 \times 10^{-4} \ll E_{J}. \nonumber
\end{eqnarray}
In other words, the states $n=10$ and $n=14$ are nearly symmetrical with
respect to the state $n=12$, which is localized in the $2$:$1$ resonance.
Therefore, the nonzero coupling $V_{2:1}$ will efficiently connect the states
$n=10$ and $n=14$.
Moreover, for the given parameters, using the definition of
$V_{r:s}(J)$ in terms of Fourier coefficient
$V_{1}$, the strength of the resonance
is estimated as $V_{2:1}(J_{2:1})\approx 0.01464$, clearly a fairly strong resonance.
Consequently, the situation in Fig.~\ref{FIG11} is
a perfect example where RAT can play a crucial role in the dissociation dynamics. 
Indeed quantum computations (not shown here) show that there is a Rabi-type cycling 
of the probabilities between the three Morse states. 
I now turn to the issue of {\em selectively} controlling the dissociation dynamics of 
the initial Morse eigenstate $n=10$
by creating local barriers in the phase space shown
in Fig.~\ref{FIG11}, bearing in mind the possibility of quantum dynamical tunneling
interfering with the control process. 

\subsubsection{Creating a local phase space barrier}

If the classical mechanism of chaotic diffusion leading to dissociation holds in the
quantum domain as well then a simple way of controlling the dissociation
is to create a local phase space barrier between the state of interest and the
chaotic region. In a recent work\cite{chandre}, Huang, Chandre and Uzer provided the theory
for recreating local phase space barriers for time-dependent systems and showed that such barriers indeed
suppress the ionization of a driven atomic system. However, Huang {\it et al.}
were only concerned with the classical ionization process. Thus, potential
complications due to dynamical tunneling were not addressed in their study. The driven
Morse system studied here presents an ideal system to understand the
interplay of quantum and classical dissociation mechanisms. In what follows,
I provide a brief introduction to the methodology with an explicit expression
for the classical control field needed to recreate an invariant KAM barrier,
preferably an invariant torus with sufficiently irrational frequency $\Omega_{r}$.

To start with, the nonautonomous Hamiltonian is mapped into an autonomous one
by considering $(t ({\rm mod} 2\pi),E)$ as an additional angle-action pair.
Denoting the action and angle variables by ${\bf A} \equiv (J,E)$ and
${\bm \theta} \equiv (\theta,t)$, 
the original driven system Hamiltonian (see Eq.~\ref{hamact2}) can be expressed as
\begin{equation}
H({\bf A},{\bm \theta})=H_{0}({\bf A}) - \epsilon V({\bf A},{\bm \theta}),
\end{equation}
with $V({\bf A},{\bm \theta})\equiv v(J,\theta;t)$.
Note that for a fixed driving field strength $\lambda_{1}$
and the value of $d_{1}$ corresponding to
a diatomic molecule, $\epsilon \equiv \lambda_{1}d_{1}$ is also fixed.
Moreover, for physically meaningful values of $d_{1}$ for most diatoms
and typical field strengths far below the ionization threshold one always has
$\epsilon \ll 1$.
In the absence of the driving field
($\epsilon=0$), the zeroth-order
Hamiltonian is integrable and the phase space is foliated
with invariant tori labeled by the action
${\bf A}$ corresponding to the
frequency ${\bm \omega} \equiv \partial H_{0}/\partial {\bf A}
=(\Omega_{0},\omega_{F})$.
However, in the presence of the driving field ($\epsilon \neq 0$)
the field-matter interaction renders the
system nonintegrable with a mixed regular-chaotic phase space.
More specifically, for field strengths near or above a critical
value $\epsilon_{c}$ one generally observes a large scale destruction of
the field-free invariant tori leading to significant chaos and hence the
onset of dissociation. The critical value $\epsilon_{c}$ itself is clearly
dependent on the specific molecule and the initial state of interest.
The aim of the local control method
is to rebuild a nonresonant torus ${\bf A}_{0}=(J_{0},0)$,
${\bf k} \cdot {\bm \omega} \neq 0$
with integer ${\bf k}$, which has been destroyed
due to the interaction with the field.
Assuming that the destruction of ${\bf A}_{0}$ is responsible
for the significant dissociation observed for some initial state of interest, the hope is
that locally recreating the ${\bf A}_{0}$ will suppress the dissociation
{\it i.e.,} ${\bf A}_{0}$ acts as a local barrier to dissociation.
Ideally, one would like to recreate the local barrier by using a second field (appropriately
called as the control field)
which is much weaker and distinct from the primary driving field.

Following Huang\cite{chandre} {\it et al.}
such a control field $f({\bm \theta})$
can be analytically derived and has the form
\begin{eqnarray}
f({\bm \theta})= -H({\bf A}_{0}-\partial_{\bm \theta}\Gamma b({\bm \theta}),{\bm \theta}),
\label{controlterm}
\end{eqnarray}
where $b({\bm \theta}) \equiv H({\bf A}_{0},{\bm \theta}) =
\sum_{{\bf k}} b_{{\bf k}} e^{i {\bf k} \cdot {\bm \theta}}$
and $\Gamma$ being a linear operator defined by
\begin{equation}
\Gamma b({\bm \theta}) \equiv \sum_{{\bf k} \cdot {\bm \omega} \neq 0}
   \frac{b_{\bf k}}{i {\bf k} \cdot {\bm \omega}} e^{i {\bf k} \cdot {\bm \theta}}.
\label{linop}
\end{equation}
The classical control Hamiltonian can now be written down as
\begin{eqnarray}
H_{c}({\bf A},{\bm \theta})&=&H({\bf A},{\bm \theta})+f({\bm \theta}) \nonumber \\
&\equiv& H_{0}({\bf A}) - \epsilon V({\bf A},{\bm \theta}) + f({\bm \theta}).
\end{eqnarray}
In case of the driven Morse system the control field can be obtained analytically
and to leading order is given by
\begin{equation}
H_{c}(J,\theta;t) \approx  H_{0}(J)-\epsilon v(J,\theta;t)+\epsilon^{2}g_{a}(\theta,t),
\label{hamcont}
\end{equation}
where 
\begin{equation}
g_{a}(\theta,t) =
 \frac{{\omega_0}^2}{4D_0}{(\partial_{\theta} \Gamma b)}^2
-2 V_{01} (\partial_{\theta} \Gamma b)
\cos(\omega_{F}t)
- (\partial_{\theta} \Gamma b) \zeta(J,\theta;t),
\label{control2}
\end{equation}
and it can be shown that
\begin{eqnarray}
V_{01} &=& \frac{{\omega_{0}}^2 }{8\alpha \Omega_{r} D_{0}}
\left(\frac{2\omega_{0}+\Omega_{r}}{\omega_{0}+\Omega_{r}}\right), \nonumber \\
V_{n1} &=& (-1)^{n+1}\left(\frac{{\omega_{0}}^3 }{2\alpha D_0}\right)
\frac{{(\omega_{0}-\Omega_{r})}^{{\frac{n}{2}}-1}}
{{(\omega_{0}+\Omega_{r})}^{\frac{n}{2}+1}}, \\
\zeta(J,\theta;t) &=& \sum_{n=1}^{\infty} V_{n1}(J_{r})
[(\cos (n\theta+{\omega}_{F} t)+\cos (n\theta-{\omega}_{F} t)]. \nonumber
\end{eqnarray}
I skip the somewhat tedious derivation of the result above and refer to
the original literature\cite{chandre} as well as a recent thesis\cite{asthesis} for details. 
Note that in the above $\Omega_{r}$ is the frequency of the invariant torus
that is to be recreated corresponding to the unperturbed action
\begin{equation}
J_{r}=(\omega_{0}-\Omega_{r})\frac{2 D_0}{\omega_{0}^2},
\end{equation}
and to $O(\epsilon)$ is located at
$J(\bm \theta)=J_{r}-\epsilon \partial_{\theta}\Gamma b$, assuming the validity of
the perturbative treatment.

The leading order control field in Eq.~\ref{control2} is typically weaker than the
driving field and has been shown\cite{asthesis} to be quite effective in off-resonant cases
in suppressing the classical dissociation. However, in order to study the effect of the
control field on the quantum dissociation probabilities, it is necessary to make
some simplifications. One of the main reasons for employing the simplified control fields
via the procedure given below has to do with the fact that the classical action-angle
variables do not have a direct quantum counterpart\cite{carru68}. Essentially, the dominant
Fourier modes $F_{k_{1},k_{2}}$ of Eq.~\ref{control2} are identified and
one performs the mapping
\begin{equation}
F_{k_{1},k_{2}} \cos(k_{1} \theta + k_{2}\omega_{F}t) \rightarrow
\lambda_{2}(k_{1},k_{2}) \cos(k_{2}\omega_{F}t),
\end{equation}
yielding the simplified control Hamiltonian
\begin{equation}
H_{c} = H(J,\theta;t) + \mu(x) \lambda_{2}(k_{1},k_{2}) \cos(k_{2} \omega_{F}t).\label{qcham}
\end{equation}
If more than one dominant Fourier modes are present then
they will appear as additional terms in Equation~\ref{qcham}.
Note that the above simplified form is equivalent to assuming that the control field
is polychromatic in nature, which need not be true in general. 
Nevertheless, a qualitative understanding of the role of the various Fourier modes
towards local phase space control is still obtained. More importantly, and as shown next, in the 
on-resonant case of interest here, the simplified control field already suggests
the central role played by RAT.

\subsubsection{RAT spoils local phase space control}

\begin{center}
\begin{figure}[t]
\includegraphics[height=100mm,width=140mm]{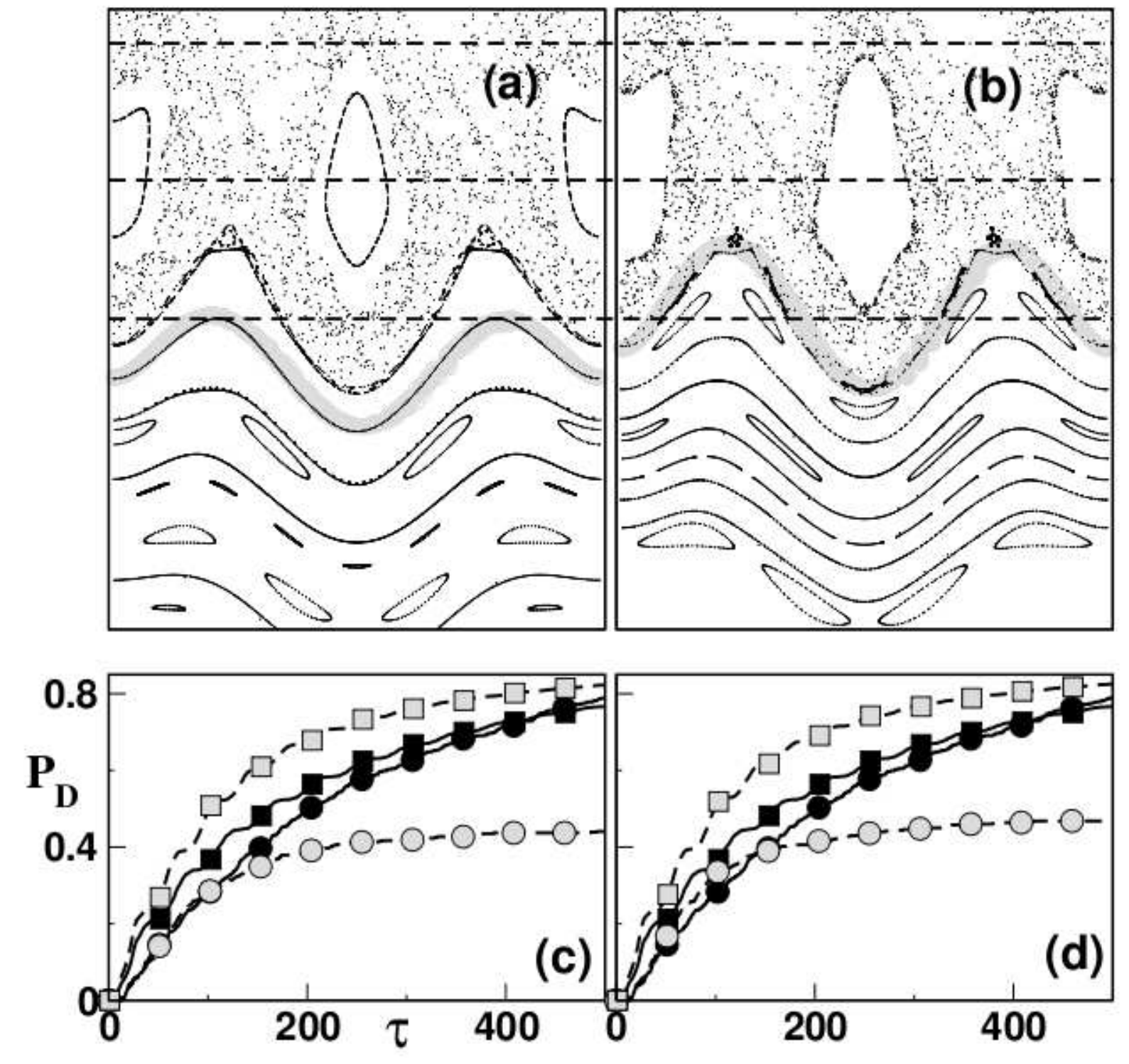}
\caption{Phase space for the driven Morse system in the presence of the
simplified classical control fields (cf. Eq.~\ref{qcham} with $\lambda_{2} \approx 0.01$)
designed to recreate specific KAM barriers
(a) $\omega_{F}/\Omega_{r}=1+\gamma_{g}^{-1}$ and
(b) $\omega_{F}/\Omega_{r}=\sqrt{3}$. Note that in both cases
the dominant $F_{3,-2}$ Fourier amplitude of the leading order control
field of Eq.~\ref{control2} have been utilized and the desired KAM barriers
are clearly seen (thick gray). The effect of the barriers seen in (a) and (b)
on the classical (gray circles) and quantum (gray squares) dissociation probabilities are
shown in (c) and (d) respectively. The uncontrolled results are indicated
by the corresponding black symbols. Interstingly, the classical dissociation is
suppressed but the quantum dissociation is slightly enhanced.}
\label{FIG12}
\end{figure}
\end{center}

In Fig.~\ref{FIG12} a summary of the efforts to control the dissociation of the
initial $n=10$ state is shown. Specifically, Fig.~\ref{FIG12}(a) and (b) show
the phase spaces where two different KAM barriers, $\omega_{F}/\Omega_{r}=1+\gamma_{g}^{-1}$
and $\omega_{F}/\Omega_{r}=\sqrt{3}$ respectively are recreated. The control Hamiltonian in both instances,
in cartesian variables, has the following form
\begin{equation}
H_{c} = H_{0}(x,p)-\lambda_{1} \mu(x) \cos(\omega_{F}t) + \lambda_{2} \mu(x) \cos(2\omega_{F}t),
\end{equation}
with $\lambda_{2} \approx 0.01 \equiv 3$ TW/cm$^{-2}$. Thus, as desired, the control
field strengths are an order of magnitude smaller than the driving field strength.
The simplified control Hamiltonian reflects the dominance of the $F_{3,-2}$ Fourier mode
of the leading order control field in Eq.~\ref{control2} and obtained using the method
outlined before. Also, note that the control field comes with a relative phase $\phi=\pi$
with respect to the driving field. As expected, from the line of thinking presented in the
previous section, Fig.~\ref{FIG12}(c) and (d) show that
the KAM barriers indeed suppress the classical dissociation significantly.
{\em However, the quantum dissociation in both cases increases slightly!}
Clearly, the recreated KAM barriers are ineffective and suggests that
the quantum dissociation mechanism is somehow bypassing the KAM barriers.

A clue to the surprising quantum results comes from comparing the phase spaces
in Fig.~\ref{FIG11} and Fig.~\ref{FIG12} which show the uncontrolled and controlled
cases respectively. Although, the KAM barriers seem to have reduced the extent of stochasticity,
the $2$:$1$ resonance is intact and appears to occupy slightly larger area in the phase space.
Thus, this certainly indicates that a significant amount of the quantum dissociation
is occuring due to the RAT mechanism involving the three Morse states $n=10,12$, and $14$
as discussed before. In particular, the $n=12$ state must still be actively providing a route
to couple the initial state to the chaotic region via RAT.
How can one test the veracity of such an explanation? One way is to scale the Planck constant
down from $\hbar=1$ and monitor the quantum dissociation process. Reduced $\hbar$
implies that the $2$:$1$ island can support several states as well as the
fact that other higher order resonances now become relevant to the RAT mechanism.
Such a study is not presented here but one would anticipate that the dissociation
mechanism can be understood based on the theory of RAT that already exists (See contributions
by Schlagheck {\it et al.} and B\"{a}cker {\it et al.}, for example).
Another way is to directly interfere locally
with the resonance and see if the quantum dissociation is actually suppressed.
Locally interfering with a specific phase space structure, keeping the gross
features unchanged, is not necessarily a straightforward approach. However, the
tools in the previous section allow for such local interference and I present the
results below. 

\begin{figure}[t]
\begin{center}
\includegraphics[height=100mm,width=140mm]{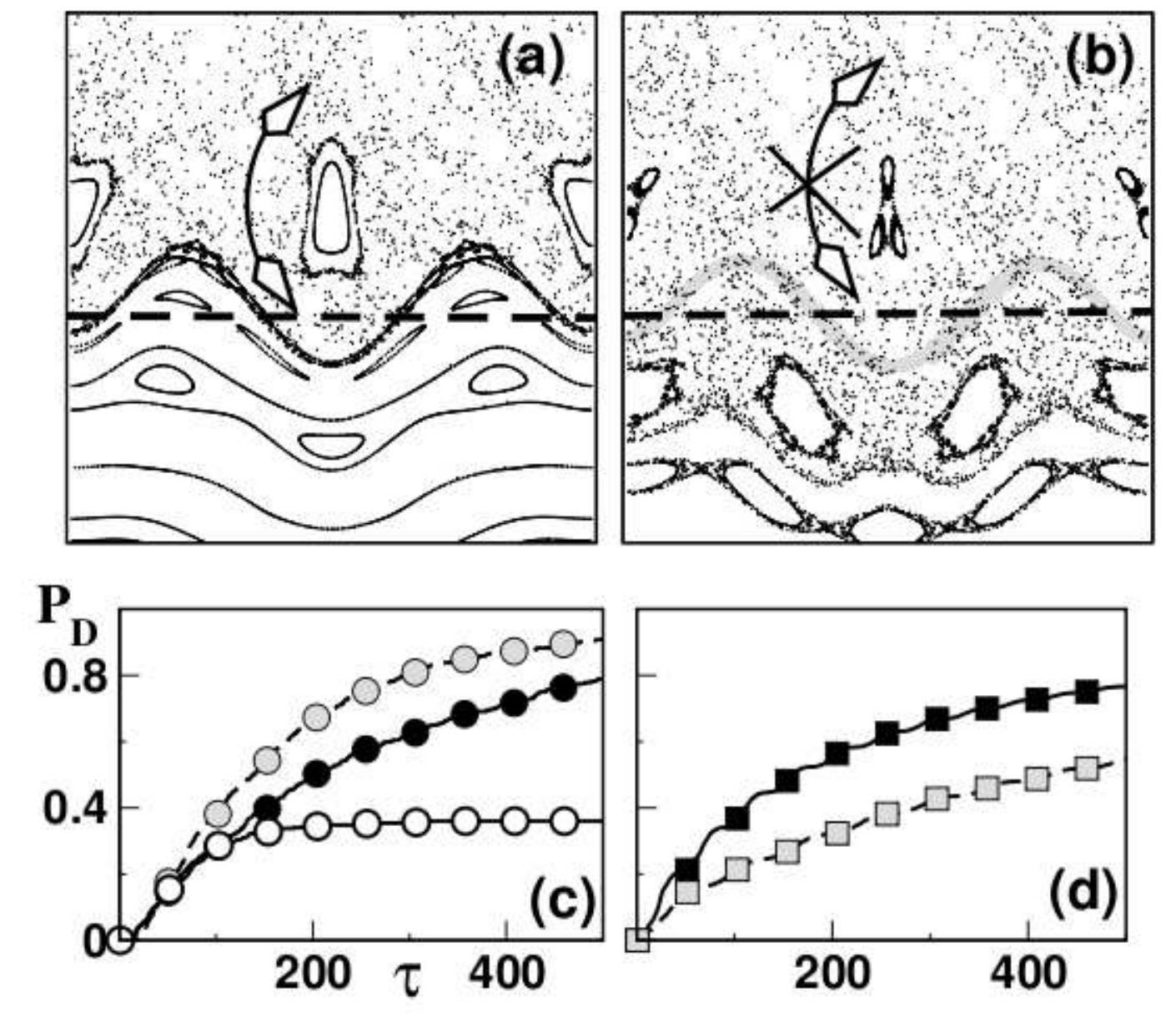}
\caption{Phase space for the driven Morse system in the presence of the
classical control fields designed to recreate the $\omega_{F}/\Omega_{r}=\sqrt{3}$
KAM barrier. In (a) two dominant Fourier modes $F_{3,-2}$ and $F_{4,-2}$ of the leading order control
term in Eq.~\ref{control2} are retained. In (b) the simplified control field
as in Eq.~\ref{qcham} is used with an effective field strength estimated using
$F_{3,-2}$ and $F_{4,-2}$. Notice that the desired KAM barrier is recreated in (a)
but not in (b) (thick gray line indicates the expected location). In (c) the classical dissociation probabilities
of the uncontrolled (black circles), control using field (a) (open circles),
and control using field (b) (gray circles) are shown. The quantum
results are shown in (d) for the uncontrolled (black squares) and control using
field (b) (gray squares) are shown. See text for discussion.}
\label{FIG13}
\end{center}
\end{figure}

The control field in Eq.~\ref{control2} corresponding to the case of Fig.~\ref{FIG12}(b)
{\it i.e.,} recreating the $\omega_{F}/\Omega_{r}=\sqrt{3}$ KAM barrier,
turns out to be well approximated by
\begin{equation}
H_{c}(J,\theta,t) \approx H(J,\theta,t)
+\sum_{n=3,4}^{} F_{n,-2} \cos(n\theta-2\omega_{F}t),
\label{appcontham}
\end{equation}
The above Hamiltonian comes about due to the fact\cite{asthesis} that two Fourier modes
$F_{3,-2}$ and $F_{4,-2}$ are significant in this case.
Note that the specific KAM barrier of interest has a frequency between that
of the $1+\gamma_{g}^{-1}$ cantorus (around which the initial state is localized, cf. 
Fig.~\ref{FIG11}) and the $2$:$1$ nonlinear resonance. From a perturbative
viewpoint, creation of this KAM barrier using Eq.~\ref{appcontham} is not expected to be easy
due to the proximity to the $2$:$1$ resonance and the fact that the $F_{4,-2}$ Fourier component
is nothing but the $2$:$1$ resonance. Nevertheless, Fig.~\ref{FIG13}(a) shows that 
the specific KAM barrier is restored
and, compared to Fig.~\ref{FIG11}, the controlled phase space does exhibit reduced
amount of chaos. Consisitently, Fig.~\ref{FIG13}(c) shows that the classical dissociation
is suppressed by nearly a factor of two as in the case shown in Fig.~\ref{FIG12}(b) wherein
only the $F_{3,-2}$ component of the control Hamiltonian was retained.
As mentioned earlier, in order to calculate the quantum dissociation probability
the control Hamiltonian in Eq.~\ref{appcontham} needs to be mapped into
a from as in Eq.~\ref{qcham}. Since two Fourier modes need to be taken into account, the
effective control field strength is given by
\begin{equation}
\lambda_{2} = \frac{F_{3,-2}}{V_{3}(J_{r})} + \frac{F_{4,-2}}{V_{4}(J_{r})}.
\label{effl2}
\end{equation}
Such a procedure yields $\lambda_{2} \approx -0.016$ and thus the control
field, still less intense than the primary field, comes with a relative
phase of zero. Interestingly, as shown in Fig.~\ref{FIG13}(b), the resulting
simplified control Hamiltonian fails to create the desired barrier. Moreover,
the phase space also exhibits increased stochasticity and as a consequence
the classical dissociation is enhanced (cf. Fig.~\ref{FIG13}(c)). However,
Fig.~\ref{FIG13}(b) reveals an interesting feature - the $2$:$1$ resonance is
severly perturbed. This perturbation is a consequence of including the
$F_{4,-2}$ Fourier component into the effective control Hamiltonian.
{\em The key result, however, is shown in Fig.~\ref{FIG13}(d) where one observes that
the quantum dissociation probability is reduced significantly}.
It seems like the quantum dynamics feels the barrier when there is none!
The surprising and counterintutive results summarized in Fig.~\ref{FIG12}
and Fig.~\ref{FIG13} can be rationalized by a single phenomenon - dynamical (resonance assisted
in this case) tunneling. The main clue comes from the observation that quantum suppression
happens as soon as the $2$:$1$ resonance is perturbed. 

Although the results here are shown with
a specific example, the phenomenon is general. Indeed computations (not published) for different
sets of parameters have supported the viewpoint expressed above. Interestingly, the work of
Huang {\it et al.} focused on classical suppression of ionization by
creating local phase space barriers in case of the
driven one-dimensional hydrogen atom\cite{chandre}. Around the same time Brodier {\it et al.} highlighted\cite{wseb06}
the importance of the RAT mechanism in order to obtain accurate decay lifetimes of
localized wavepackets in the same system. Work is currently underway to see if the
conclusions made in this section hold in the driven atomic system as well {\it i.e.,}
whether the attempt to suppress the ionization by creating local phase space barriers is
foiled by the phenomenon of RAT.

\subsection{Summary and future outlook}

The two examples discussed in this work illustrate a key point - {\em Dynamical tunneling
plays a nontrivial role in the process of quantum control.} In the first example of 
the driven quartic double well it is clear that bichromatic control fails in regimes where
chaos-assisted tunneling is important. In the second example of a driven Morse oscillator
it is apparent that efforts to control by building phase space KAM barriers fail
when resonance-assisted tunneling is possible. In both instances the competition
between classical and quantum mechanisms is brought to the forefront. Although the two examples shown here
represent the failure of specific control schemes due to dynamical tunneling, one must not
take this to be a general conclusion. It is quite possible that some other control schemes
might owe their efficiency to the phenonmenon of dynamical tunneling itself. 
Further classical-quantum correspondence studies on control with more general driving fields are required 
in order to confirm (or refute) the conclusions presented in this chapter.

At the same time
the two examples presented here are certainly not the the last word; establishing the role
of dynamical tunneling in quantum coherent control requires one to step into the murky
world of three or more degrees of freedom systems\cite{acp130}. I mention two model systems, currently
being studied in our group, in order to stress upon some of the key issues that might crop up in
such high dimensional systems. 
For example, two coupled Morse oscillators which
are driven by a monochromatic field already presents a number of challenges
both from the technical as well as conceptual viewpoints. The technical challenge arises due
to the fact that dimensionality constraints do not allow one to visualize the global phase space structures as easily as done in
this chapter. One approach is to use the method of
local frequency analysis\cite{lfa} to construct the Arnol'd web {\it i.e.,} the network of nonlinear
resonances that regulate the multidimensional phase space transport. In a previous
work\cite{Kes051}, involving a time-independent Hamiltonian system, the utility of such an approach
and the validity of the RAT mechanism has been established. However, ``lifting" quantum
dynamics onto the Arnol'd web is an intriguing possibility which is still an open issue.
On the conceptual side there are several issues with multidimensional systems. I mention
a few of them here. Firstly, even at the classical level one has the possibility of 
transport like Arnol'd diffusion\cite{lichtlieb,pr79-ch-chap3} which is genuinely a three or more DoF effect
and has no counterpart in systems with less than three DoFs. Note that Arnol'd diffusion 
is typically a very long time process and is notoriously
difficult to observe in realistic physical systems\cite{ardifnot}.
Moreover, arguments can be made for the irrelevance of Arnol'd diffusion (or some
similar process) in quantum systems due to the finiteness of the Planck constant.
Secondly, an interesting competition occurs in systems such as the driven coupled Morse
oscillators. Even in the absence of the field the dynamics is nonintegrable and
one can be in a regime where the modes are exchanging energy but none of the modes
gain enough energy to dissociate. 
On the other hand, in the absence of mode-mode coupling, a weak enough field
can excite the system without leading to dissociation. However,
in the presence of such a weak field and the mode-mode coupling one can have
significant dissociation of a specific vibrational mode. Clearly, there is nontrivial
competition between transport due to mode-mode resonances, field-mode resonances and
the chaotic regions\cite{asthesis}. Selective control of such driven coupled systems is an active
research area\cite{cohcontrol} today and the lessons learnt from the two examples suggest
that dynamical tunneling in one form or another can play a central role.
In the context of studying the potential competition between Arnol'd diffusion
and dynamical tunneling, I should mention the driven coupled quartic oscillator
system with the Hamiltonian:
\begin{equation}
H(x,y,p_{x},p_{y};t) = \frac{1}{2}(p_{x}^{2}+p_{y}^{2})+\frac{1}{4}(x^{4}+y^{4}) - \mu x^{2}y^{2} 
                       -x f_{0}(\cos \Omega_{1}t + \cos \Omega_{2} t).
\end{equation}
In the absence of the field $f_{0}=0$ the Hamiltonian reduces to the case originally
studied by Tomsovic, Bohigas and Ullmo wherein the existence of chaos-assisted tunneling was
established in exquisite detail\cite{btu93,tu94}. At the same time, with $f_{0} \neq 0$ the system represents
one of the few examples for which the phenomenon of Arnol'd diffusion has been investigated
over a number of years. Nearly a decade ago, Demikhovskii, Izrailev, and Malyshev studied\cite{dim02} a variant of
the above Hamiltonian to uncover the fingerprint of Arnol'd diffusion on the quantum eigenstates
and dynamics. An interesting question, amongst many others, is this: {\em Will the fluctuations
in the chaos-assisted tunneling splittings for $f_{0}=0$, observed for varying $\hbar$, survive
in the presence of the field?}

A related topic which I have not touched upon in this article has to do with the
control of IVR using weak external fields. Based on the insights gained from studies done
until now on field-free IVR (see also the contribution by Leitner in this volume), it is natural to
expect that dynamical tunneling could play spoilsport for certain class of initial
states that are prepared experimentally. The issue, however, is far more subtle
and more studies in this direction might shed light on the mechanism by which
quantum optimal control methods work. For instance, recent proposals on quantum control 
by Takami and Fujisaki\cite{tf07} and the so called ``quantized Ulam conjecture" by
Gruebele and Wolynes\cite{gwulam} take advantage (implicitly) of the system having a completely chaotic 
phase space. Dynamical tunneling is not an issue in such cases. However, in more generic instances
of systems with mixed regular-chaotic phase space, a quantitative and qualitative
understanding of dynamical tunneling becomes imperative. Given the level of detail at which one is now capable of
studying the clasically forbidden processes, as reflected by the varied contributions
in this volume, I expect exciting progress in this direction.

\section{Acknowledgements}

I think that this is an appropriate forum to acknowledge the genesis of the current
and past research of mine on dynamical tunneling. In this context I am grateful to Greg Ezra,
whose first suggestion for my postdoc work came in the form of a list of some of the key papers on dynamical
tunneling. We never got around to work on dynamical tunneling {\it per se} but those references 
came in handy nearly a decade later. It is also a real pleasure to thank Peter Schlagheck
for several inspiring discussions on tunneling in general and dynamical tunneling in particular.

\end{document}